\documentclass[11pt, a4paper]{amsart}
\usepackage[dvips]{epsfig}
\usepackage{amsmath}
\usepackage{amssymb}
\usepackage{amsfonts}
\usepackage{amsthm}
\usepackage{amsbsy}
\usepackage{amsgen}
\usepackage{amscd}
\usepackage{amsopn}
\usepackage{amstext}
\usepackage{amsxtra}

\newtheorem{theorem}{Theorem}[section]
\newtheorem{proposition}[theorem]{Proposition}
\newtheorem{lemma}[theorem]{Lemma}
\newtheorem{corollary}[theorem]{Corollary}
\newtheorem{defn}[theorem]{Definition}

\theoremstyle{definition}

\numberwithin{equation}{section}

\newcommand {\Z} {\mathbb{Z}}

\newcommand {\R} {\mathbb{R}}
\newcommand {\T} {\mathbb{T}}

\newcommand {\E} {\mathbb{E}}
\newcommand {\Q} {\mathbb{Q}}

\newcommand {\B} {\mathcal {B}}

\newcommand {\PP} {\mathcal P}

\newcommand{\meas}{\operatorname{meas}}
\newcommand{\Px}{\PP_x}
\newcommand{\Pxy}{\PP_{x,y}}
\newcommand{\Pxz}{\PP_{x,x+z}}
\newcommand{\setoutx}{Sing_x^{\rm out}}
\newcommand{\setinx}{Sing_x^{\rm in}}
\newcommand{\Lpm}{\Lambda/\pm}

\newcommand{\Ndim}{\mathcal N}
\newcommand{\eigenvalue}{E}
\newcommand{\eigenspace}{\mathcal{E}}
\newcommand{\leray}{\mathcal L}
\newcommand{\TT}{{\mathbb T}}
\newcommand{\var}{\operatorname{Var}}

\begin{document}

\title[Leray measure for nodal sets of eigenfunctions on the torus]
{The Leray measure of nodal sets for random eigenfunctions on the torus }

\author{Ferenc Oravecz, Ze\'ev Rudnick and Igor Wigman}
\address{Alfr\'ed R\'enyi Institute of Mathematics,  Hungarian Academy
of Sciences, Re\'altanoda utca 13-15, H-1053 Budapest, Hungary} 
\email{oravecz@renyi.hu}
\address{School of Mathematical Sciences, Tel Aviv University,
Tel Aviv 69978, Israel}
\email{rudnick@post.tau.ac.il}
\address{Centre de recherches math\'ematiques (CRM), 
Universit\'e de Montr\'eal
C.P. 6128, succ. centre-ville
Montr\'eal, Qu\'ebec H3C 3J7, Canada} 
\email{wigman@crm.umontreal.ca} 
\date{January 2, 2007} 
\begin{abstract}
We study  nodal sets for typical eigenfunctions of the Laplacian on
the standard torus in $d\geq 2$ dimensions.
Making use of the multiplicities in the spectrum of the
Laplacian, we put a Gaussian  measure on the eigenspaces and use it to
average over the eigenspace.
We consider a sequence of eigenvalues with growing  multiplicity
$\Ndim\to\infty$.

The quantity that we study is the Leray, or microcanonical, measure of
the nodal set. We show that the expected value of the Leray measure of
an eigenfunction is constant, equal to $1/\sqrt{2\pi}$.
Our main result is that the variance of Leray measure is
asymptotically  $1/4\pi\Ndim$, as $\Ndim\to \infty$,
at  least in dimensions $d=2$ and $d\geq 5$.
\end{abstract}
\maketitle

\tableofcontents

\section{Introduction}

\subsection{Background}

The {\em nodal  set} of a  function is the set of points where the
function  vanishes.
In this paper we study the nodal sets of eigenfunctions of the
Laplacian $\Delta =\sum_{j=1}^d \frac{\partial^2}{\partial x_j^2}$ on
the (standard) flat torus $\R^d/\Z^d$, $d\geq 2$.

Of course we have the simple  eigenfunctions such as
$\cos(2\pi(mx+ny))$  or $\sin (2\pi mx) \sin(2\pi ny)$
with corresponding Laplace eigenvalue $4\pi^2(m^2+n^2)$,
for which the nodal set have a very simple structure.
However, on the standard torus such eigenfunctions are atypical,
because the eigenvalues on the torus always have multiplicities.
The dimension $\Ndim=\Ndim(\eigenvalue)$ of an eigenspace corresponding
to eigenvalue $4\pi^2\eigenvalue$ is
the number of integer vectors $\lambda\in \Z^d$ so that
$|\lambda|^2=\eigenvalue$. In dimension $d\geq5$ this grows as
$\eigenvalue\to\infty$ roughly as $\eigenvalue^{\frac d2-1}$ but has
more erratic behaviour for small $d$, particularly for $d=2$.

We wish to study the nodal sets of {\em typical} eigenfunctions. For
this we consider a {\em random} eigenfunction on the torus, that
is a random linear combination
$$
f(x)=\frac 1{\sqrt{2\Ndim}}
\sum_{\lambda\in \Z^d: |\lambda|^2=\eigenvalue}
b_\lambda \cos 2\pi i \langle \lambda, x \rangle
- c_\lambda \sin 2\pi i \langle \lambda, x \rangle
$$
with $b_\lambda,c_\lambda\sim N(0, 1)$ real  Gaussians of zero
mean and variance $1$ which are independent save for the relations
$b_{-\lambda}=b_\lambda$, $c_{-\lambda}=-c_\lambda$.

We denote by $\E(\bullet)$ the expected value of the quantity
$\bullet$ in this ensemble.
For instance, the expected amplitude of $f$ is $\E(|f(x)|^2)=1$.

\subsection{Leray measure}
The fundamental quantity that we study here is the Leray measure, or
microcanonical measure, of the nodal set of a function $f$ in our
ensemble. This is defined as
(see \cite[Chapter III]{GS1}, \cite[\S 3.3]{Palamodov})
\begin{equation}\label{eq:def of leray}
\leray(f):=\lim_{\epsilon\to 0} \frac 1{2\epsilon}
\meas \{x\in \TT: |f(x)|<\epsilon \}.
\end{equation}
and in fact we can define a measure on the nodal set by
$$
 \lim_{\epsilon\to 0} \frac 1{2\epsilon}
\int_{x:|f(x)|<\epsilon} \phi(x) dx
$$
which in statistical mechanics is the microcanonical ensemble. This
measure also appears in number theory as the ``singular integral''  in
the Hardy-Littlewood method and elsewhere, see e.g. \cite{Davenport, BR}.
We may formally write
$$
\leray(f)=\int_{\TT^d} \delta(f(x)) dx\;.
$$

As is well known, the limit \eqref{eq:def of leray} exists when $\nabla
f\neq 0$ on the nodal set, in which case
$$\leray(f) = \int_{\{x:f(x)=0\}} \frac{d\sigma(x)}{|\nabla f(x)|}
$$
where $d\sigma$ is the Riemannian hypersurface measure on the nodal
set (see \S \ref{sec:expect}).


\subsection{Results}
The expected value of $\leray(f)$
turns out to be constant (Theorem~\ref{prop:exp nod leb}):
$$
\E(\leray) = \frac 1{\sqrt{2\pi}} \;.
$$
To compare, the expected {\em volume} (or hypersurface measure) of the
nodal set  of $f$ in our ensemble is ${\mathcal I}_d
\sqrt{\eigenvalue}$ for some 
constant $\mathcal I_d$ depending only on the dimension \cite{RW}.

Our main result concerns the {\em variance} of $\leray(f)$ as $\Ndim
\to \infty$:
\begin{theorem}\label{thm:var}
In dimensions $d=2$ and $d\geq 5$, as $\Ndim \to \infty$,
$$
\var(\leray(f)) \sim \frac 1{4\pi \Ndim} \;.
$$
\end{theorem}
We refer to  \cite{RW} for estimates on the variance of the 
volume of the nodal sets.  

Concerning remainder terms, in dimension $d=2$ we show that 
$\var(\leray(f)) = 1/4\pi \Ndim  + O(1/\Ndim^2)$. In dimension
$d\geq 3$, we prove $\var(\leray(f)) = 1/4\pi \Ndim  +
O(\eigenvalue^{\frac{d-3}2+\epsilon}/\Ndim^2)$, for all $\epsilon>0$.
Thus whenever $\Ndim>\eigenvalue^{\frac{d-3}2+\delta}$ for some
$\delta>0$ (which is always valid in dimension $d\geq 5$), then we get
an asymptotic. 
In dimensions $d=3,4$ we are only able to show that the variance is
bounded by $O(1/\Ndim)$, though we believe that the conclusion of
Theorem~\ref{thm:var} holds in those cases as well.

It is somewhat surprising that the result depends only on the
 dimension of the eigenspace and not on the way the frequencies
 $\lambda$ are  distributed. In dimension $d\geq 5$, the directions
 $\lambda/|\lambda|$ of the frequencies  are uniformly
 distributed on the sphere $S^{d-1}$ \cite{Pommerenke}.
However, in  two dimensions this need not be the case
(though it holds for most values of $\eigenvalue$,
see \cite{EH, KK, FKW}).
For instance there is an infinite sequence of eigenvalues
 where the dimension of the eigenspace goes to infinity but the set of
 directions $\lambda/|\lambda| \in S^1$ tends to an average of four
equally spaced  point masses \cite{Cilleruelo}.

\subsection{Related work}
The study of nodal lines of random waves goes back to Longuet-Higgins
\cite{LH1,LH2} who computed various statistics of nodal lines for
Gaussian random waves in connection with the analysis of ocean waves.
Berry \cite{Berry 1977} suggested to model highly excited quantum
states for classically chaotic systems by using various random wave
models, and also computed fluctuations of various quantities in these
models (see e.g. \cite{Berry 2002}). See also Zelditch
\cite{ZelditchRMT}. 
The idea of averaging over a single eigenspace in the presence of
multiplicities appears in B\'erard \cite{Berard} who computed the
expected surface measure of the nodal set for eigenfunctions of the
Laplacian on spheres. Neuheisel \cite{Neuheisel} also worked on the
sphere and  studied the statistics of Leray measure. He gave
an upper bound for the variance, which we believe is not sharp.  

\subsection{About the proof of Theorem~\ref{thm:var}}
We compute the second moment $\E(\leray^2)$ by means of Gaussian
integration as an integral over the torus
$$\E(\leray^2) = \frac 1{2\pi} \int_{\TT^d}\frac {dx}{\sqrt{1-u(x)^2}}
$$
where
$$
u(x) := \E(f(x+y) f(y))  = \frac 1{\Ndim}
\sum_{|\lambda|^2=\eigenvalue} \cos 2\pi \langle \lambda,x \rangle
$$
is the two-point function of our random process (which is translation
invariant).  This formula shows that one should single out
points $x\in \TT^d$ where $|u(x)|$ is close to $1$ (clearly
$|u(x)|\leq 1$).
We will show (see section \ref{subsec:contr sing}) that the total contribution to the
integral near such (suitably defined) ``singular'' points is bounded
by  $O(\int_{\TT^d}u(x)^4dx)$.

Outside of these ``singular'' points,
we may expand in a Taylor series
$(1-u^2)^{-1/2} = 1+\frac 12 u^2 +O(u^4)$.
The constant term $1$ corresponds to the square of the
expectation and thus we will get
$$
\var(\leray) = \frac 1{4\pi}\int_{\TT^d} u(x)^2dx +
O\bigg(\int_{\TT^d}u(x)^4 dx\bigg) \;.
$$

The second moment of $u$ is immediately seen to equal
$\int_{\TT^d}u(x)^2dx=1/\Ndim$, and it is easily seen
that the fourth moment of $u$ is at most $1/\Ndim$. Thus we get an
upper bound $\var(\leray)=O(1/\Ndim)$ (in any dimension $d\geq 2$).
To obtain  Theorem~\ref{thm:var} one needs to show
that the fourth moment of $u$ is negligible relative to $1/\Ndim$.
In dimension $d=2$ we have $\int_{\TT^d}u(x)^4dx\ll 1/\Ndim^2$
by a geometric argument due to Zygmund \cite{Zygmund}. 
In dimension  $d\ge3$, we can show that 
\begin{equation}\label{upper bd for K4}
\int_{\TT^d} u(x)^4dx \ll_\epsilon
\frac{\eigenvalue^{\frac{d-3}2+\epsilon}}{\Ndim^2},\quad \forall \epsilon>0
\end{equation}
which in dimension $d\geq 5$ suffices because
$\Ndim\approx \eigenvalue^{\frac d2-1}$ and so we get a bound of
$1/\Ndim\eigenvalue^{1/2-\epsilon}$.

Alternatively, note that $u(x)$ is itself an eigenfunction of
the Laplacian and we want a  bound on its $L^4$-norm relative
to its $L^2$-norm.
In dimension $d\geq 5$  a bound
(valid for any Riemannian manifold) due to
Sogge \cite{Sogge} suffices here. A stronger bound for the torus, due
to Bourgain \cite{Bourgain}, will improve \eqref{upper bd for K4}
for $d\geq 7$.

\subsection{Acknowledgements} We thank Misha Sodin for several helpful
discussions. This work was supported by the Israel Science
Foundation (grant  No. 925/06). In addition, I.W. was partly supported by
SFB 701: Spectral Structures and Topological Methods in Mathematics, 
(Bielefeld University). 

\section{Random eigenfunctions on the torus}

\subsection{The basic setup}
We wish to consider eigenfunctions of the Laplacian on the standard
flat torus:
$$\Delta\psi+4\pi^2\eigenvalue\psi=0\;.$$
These can be written as linear combinations of the basic
exponentials $e^{2\pi i \langle \lambda,x \rangle}$,
with $\lambda\in \Z^d$, $|\lambda|^2=\eigenvalue$.
The dimension $\Ndim$ of the corresponding eigenspace is simply the number
of ways of expressing $\eigenvalue$ as a sum of $d$ integer squares.
For $d\geq 5$ this grows roughly as  $\eigenvalue^{d/2-1}$ as
$\eigenvalue\to\infty$. For $d\leq 4$ the  dimension of the eigenspace
need not grow with $\eigenvalue$. In the extreme case $d=2$,
$\Ndim$ is given in terms of the prime
decomposition of $\eigenvalue$ as follows:
If $\eigenvalue=2^\alpha\prod_j p_j^{\beta_j}\prod_k
q_k^{2\gamma_k}$  where $p_j\equiv 1 \mod 4$ and $q_k\equiv 3\mod 4$
are odd primes, $\alpha,\beta_j,\gamma_k\geq$ are integers,
then  $\Ndim = 4\prod_j(\beta_j+1)$, and otherwise $\eigenvalue$ is not
a sum of two squares and $\Ndim=0$. On average (over integers which are
sums of two squares) the dimension is $const\cdot \sqrt{\log \eigenvalue}$.

For some of our initial work, throughout sections
\S~\ref{sec:expect}, \ref{sec:variance1}
we will work in greater generality and
instead of eigenspaces we will consider linear spaces
$\eigenspace=\eigenspace(\Lambda)$ spanned by
certain sets of exponentials $e^{2\pi i\langle \lambda,x\rangle}$ with
$\lambda\in \Lambda\subset \Z^d$.
We take into account the reflection symmetries of the torus by
assuming that the frequency set $\Lambda$ is invariant under the group
of signed permutations $W_d = \{\pm 1\}^d \times S_d$, consisting of
coordinate permutations and sign-change of any coordinate,
e.g. $(\lambda_1,\lambda_2)\mapsto (-\lambda_1,\lambda_2)$ (for $d=2$).
We say that a non-empty subset $\Lambda\subset \Z^d$ is ``symmetric'' if
it is invariant under $W_d$, that is invariant under permutations of
the coordinates and changing sign of each coordinate,
and that $0\notin \Lambda$.

The dimension $\Ndim = \dim \eigenspace$ is the number of the frequencies in
$\Lambda$. Since $\Lambda$ is symmetric and does not contain $0$,
$\Ndim$ is even. We write $\Lambda/\pm$ to denote representatives of
the equivalence class of $\Lambda$ under $\lambda\mapsto -\lambda$.

\begin{lemma}
\label{lem:Lambda spans Rd}
Any set $\Lambda$ satisfying the symmetry
conditions (i.e. invariant w.r.t. coordinate permutations and sign
changes), spans $\R^d$.
\end{lemma}
\begin{proof}
Otherwise we have a nontrivial linear relation
\begin{equation}
\label{eq:lin rel lam} \sum\limits_{l=1}^{d} c_i \lambda_i = 0,
\end{equation}
valid for all $\lambda\in\Lambda$. Since $\Lambda$ is invariant
under permutations, we may assume $\lambda_1\ne 0$. Substituting
$\lambda$ and $\lambda'=(-\lambda_1',\,\lambda_2,\,\ldots,\,
\lambda_d)$ and subtracting the equations we obtain $2 \lambda_1
c_1 = 0$, which implies $c_1=0$. Repeating the argument for all
$c_i$, we get a contradiction.
\end{proof}

As a consequence of this lemma, we see that the set $L_\Lambda$ of
integer linear combinations of elements of $\Lambda\subseteq \Z^d$ is
a sublattice of full rank, and hence its dual
$$L_\Lambda^*=\{v\in \R^d: \langle \lambda, v\rangle \in \Z, \forall
\lambda \in \Lambda\}$$
is also a lattice in $\R^d$ (containing $\Z^d$).


\subsection{A non-degeneracy condition}  
Assume that the set of frequencies $\Lambda$, which is assumed to be
``symmetric'', further satisfies the following ``non-degeneracy'' condition:
\begin{equation}\label{condition 1a}
\exists \lambda \in \Lambda \mbox{ with } \lambda_1
\neq \pm \lambda_2 \mbox{ and } \lambda_1,\lambda_2 \neq 0 \;.
\end{equation} 
By the symmetry of the set $\Lambda$, condition~\eqref{condition 1a} is
equivalent to requiring that for every $i\neq j$, there is $\lambda\in
\Lambda$ with $\lambda_i\neq \pm \lambda_j$ and
$\lambda_i,\lambda_j\neq 0$. 

In the case of eigenfunctions of the Laplacian, where
$\Lambda=\{\lambda\in \Z^d: |\lambda|^2=\eigenvalue\}$, 
the non-degeneracy condition~\eqref{condition 1a} holds as soon as
$\Ndim = \#\Lambda$ is 
sufficiently large, in fact if $\Ndim> 3^d$. This is because any
$\lambda$ where there are no distinct indices $i\neq j$ with
$\lambda_i,\lambda_j\neq0$, $\lambda_i\neq \pm \lambda_j$ must be in
the $W_d$-orbit of a vector of the form $\lambda(j,r) = (r,r,\dots,
r,0,\dots,0)$ with the first $j$ coordinates equal to $r>0$ and the
remaining $d-j$ coordinates equal to zero, and $\eigenvalue=jr^2$ (so
$r$ is determined uniquely by $\eigenvalue$ and $0\leq j\leq d$). The
number of elements in the $W_d$-orbit of $\lambda(j,r)$ is
$\binom{d}{j}2^j$ and summing over all $0\leq j\leq d$ gives 
at most $3^d$ possibilities.

\subsection{Gaussian ensembles}\label{sec:Gaussian ensembles}
For any symmetric set of frequencies
$\Lambda\subset \Z^d$, we define an ensemble of Gaussian random
functions $f\in\eigenspace$ by
\begin{equation}
\label{eq:gen def f}
f(x)=\frac 1{\sqrt{2\Ndim}}
\sum_{\lambda\in \Lambda}
b_\lambda \cos 2\pi i \langle \lambda, x \rangle
- c_\lambda \sin 2\pi i \langle \lambda, x \rangle
\end{equation}
with $b_\lambda,c_\lambda\sim N(0, 1)$ real  Gaussians of zero
mean and variance $1$ which are independent save for the relations
$b_{-\lambda}=b_\lambda$, $c_{-\lambda}=-c_\lambda$.
Thus we can rewrite
$$
f(x)=\sqrt{\frac 2 {\Ndim}}
\sum_{\lambda\in \Lambda/\pm}
b_\lambda \cos 2\pi i \langle \lambda, x \rangle
- c_\lambda \sin 2\pi i \langle \lambda, x \rangle
$$
where now only independent random variables appear.

Alternatively, we may identify $\eigenspace\cong \R^\Ndim$ by taking
coordinates $Z=(b_{\lambda},\, c_{\lambda})_{\lambda\in\Lpm}$ and
putting the Gaussian probability measure
$$
d\mu_\Ndim(Z) = \frac 1{(2\pi)^{\Ndim/2}} \prod_{\lambda\in
\Lambda/\pm} e^{-(b_\lambda^2+c_\lambda^2)/2}db_\lambda dc_\lambda \;.
$$

We define a set $\B$ by
$$\B = \{w\in \R^d: \langle \lambda,w \rangle \in \Z\quad \forall
\lambda\in \Lambda \mbox{ or } 
 \langle \lambda,w \rangle \in \frac 12 +\Z\quad \forall
\lambda\in \Lambda \} \;. 
$$
Then clearly  $\frac 12 L_\Lambda^* \subseteq  \B\subseteq
L_\Lambda^*$ and so the projection of $\B$ on the torus
$\TT^d=\R^d/\Z^d$ is finite. 
Note that if $x-y\in \B$, then for all $f\in \eigenspace$,
$$ 
f(y)=\pm f(x),\quad \mbox{ and } \nabla f(y) = \pm \nabla f(y) \;. 
$$


For $a=(a_1,a_2) \in \R^2$, let 
$$
\Pxy^a = \{f\in \eigenspace: f(x)=a_1, f(y)=a_2 \}\;.
$$
If $x-y\notin \B$ then this is an affine hyperplane
of codimension two in $\eigenspace$.  
If $x-y\in \B$ then this is either empty or a
hyperplane of codimension one in $\eigenspace$.

We define the two-point function of our ensemble as
$$
u(x,y) = \E(f(x) f(y)) \;.
$$
A simple computation shows that $u(x,y)$ depends only on the
difference $x-y$, in fact $u(x,y)=u(x-y)$ where
$$
u(z) = \frac 1{\Ndim} \sum_{\lambda\in \Lambda}
\cos 2\pi \langle \lambda,z\rangle \;.
$$

\begin{lemma}
\label{lem:u(x)!=u(y) a.a}
$u(x)=\pm 1$ if and only if  $x\in \B$. 
\end{lemma}
\begin{proof}
If $x\in \B$ then $\cos2\pi \langle \lambda,x \rangle$ are all equal,
to either $+1$ or $-1$ and hence $u(x)=\pm 1$. 
On the other hand, 
since $|\cos2\pi \langle \lambda,x \rangle |\leq 1$, if $u(x)=\pm 1$
then all the
cosines $\cos2\pi \langle \lambda, x \rangle$ have the same value,
which is either $+1$ or $-1$, and this forces either $\langle \lambda, x
\rangle\in \Z$ for all $\lambda\in \Lambda$, 
or $\langle \lambda, x \rangle\in \frac 12 +\Z$ for all $\lambda\in
\Lambda$, that is  $x\in \B$. 
\end{proof}


\subsection{The singular set}

We  define the set of {\em singular}  functions 
to be
\begin{equation*}
Sing := \{ f\in \eigenspace :\: \exists x\in \T^d,\, f(x)=0 \text{ and }
(\nabla f)(x) = \vec{0} \}.
\end{equation*}

\begin{lemma}\label{lem:sing codim 1}
\label{lem:Sing meas 0} The set $Sing$ has codimension at least
$1$ in $\eigenspace$.
\end{lemma}
\begin{proof}
Define
\begin{equation*}
\begin{split}
\psi:\T^d\times\eigenspace \rightarrow\R^d\times\R \\
(x,\, f)\mapsto (\nabla f (x),\, f(x)),
\end{split}
\end{equation*}

Denoting $\pi_2: \T^d\times\R^{\Ndim}\rightarrow\R^{\Ndim}$ the
projection to the second factor, we have
\begin{equation*}
Sing = \pi_2 (\psi^{-1} (\{0\} \times \{0\})).
\end{equation*}
We prove that the Jacobian of $\psi$ has maximal rank everywhere,
and therefore $\psi^{-1} (\{0\} \times \{0\})$ is a smooth
manifold of codimension $d+1$. It will then follow that
$Sing\subset \R^{\Ndim}$ has codimension $\ge 1$.

The $(d+1)\times(d+\Ndim)$ Jacobian matrix is
\begin{equation*}
D\psi (x) = \left( \begin{matrix} * & -2\pi \sqrt{\frac{2}{\Ndim}} A(x) \\ *
&\sqrt{\frac{2}{\Ndim}} B(x)\end{matrix}\right),
\end{equation*}
where $A(x)$ is a $d\times \Ndim$ matrix defined by
\begin{equation*}
A(x) =   \bigg( (  \sin 2\pi \langle \lambda,x \rangle \vec \lambda
,\, \cos 2\pi \langle \lambda,x \rangle \vec \lambda)
\bigg) _{\lambda\in\Lpm},
\end{equation*}
and $B(x)$ is a $1\times \Ndim$ matrix defined by
\begin{equation*}
B(x) = \bigg( (\cos 2\pi \langle \lambda,x \rangle ,\, - \sin 2\pi \langle \lambda,x \rangle ) \bigg)
_{\lambda\in\Lpm}.
\end{equation*}
Thus we want the $(d+1)\times \Ndim$ matrix $\left(
\begin{matrix} A \\ B \end{matrix} \right)$ to have rank $d+1$.
However, ordering the vectors $\vec{\lambda}^{(j)}\in\Lambda/\pm$, it
is a product of
\begin{equation*}
\left( \begin{matrix} &\vec{0} &\vec{\lambda}^{(1)} &\vec{0}
&\vec{\lambda}^{(2)} &\ldots \\ &1 &0 &1 &0 &\dots
\end{matrix} \right),
\end{equation*}
which is of rank $d+1$ by lemma \ref{lem:Lambda spans Rd} and
\begin{equation*}
\begin{pmatrix} & \cos 2\pi \langle \lambda^{(1)},x \rangle
 &- \sin 2\pi \langle \lambda^{(1)},x \rangle
 &0 &0 &\ldots \\
&\sin 2\pi \langle \lambda^{(1)},x \rangle
& \cos 2\pi \langle \lambda^{(1)},x \rangle
 &0 &0 &\ldots \\ &0 &0
& \cos 2\pi \langle \lambda^{(2)},x \rangle
 &- \sin  2\pi \langle \lambda^{(2)},x \rangle
 &\ldots \\ &0 &0
&\sin  2\pi \langle \lambda^{(2)},x \rangle
 & \cos 2\pi \langle \lambda^{(2)},x \rangle
 &\dots \\ & & & & &\ddots
\end{pmatrix}
\end{equation*}
which is nonsingular. This immediately implies the result.
\end{proof}

The following is an immediate
\begin{corollary}
The set $Sing$ has measure zero in $\eigenspace$.
\end{corollary}

\section{The  Leray measure}

\newcommand{\eigmax}{\eigenvalue_{\max}}

We continue with our  previous setting, that is
$\Lambda\subset \Z^d$ is a symmetric, non-degenerate  set of frequencies.
We wish to define the Leray measure $\leray(f)$ for
$f\in \eigenspace$  by the limit
$$
\leray(f)=\lim_{\epsilon\to 0} \frac 1{2\epsilon}
\meas\{x: |f(x)|<\epsilon \}\;.
$$
It is well known that the limit exists for any
nonsingular $f$ (see \cite[Chapter III]{GS1}, \cite[\S 3.3]{Palamodov}),
and that in fact
$$\leray(f) = \int_{\{x:f(x)=0\}}  \frac{d\sigma(x)}{|\nabla f(x)|}
$$
where $d\sigma(x)$ is the induced hypersurface measure.

We will need to know more refined information about the approach to
the limit in the definition.
For $\epsilon>0$, set
$$
\leray_\epsilon(f):=\frac 1{2\epsilon}\meas\{x: |f(x)|<\epsilon\}\;.
$$
so that $\leray(f) = \lim_{\epsilon\to 0} \leray_\epsilon(f)$.

For $\alpha>0$,  $\beta>0$ let
\begin{equation*}
\eigenspace(\alpha,\, \beta) = \{f\in \eigenspace :\: |f(x)| \leq \alpha
\Rightarrow\ |\nabla f (x)| > \beta \} \;.
\end{equation*}
The sets $\eigenspace(\alpha,\, \beta)$ are open, and have the
monotonicity property
$$
\alpha_1 >\alpha_2 \Rightarrow \eigenspace(\alpha_1,\beta) \subseteq
\eigenspace(\alpha_2,\beta)
$$
 and
$$
\beta_1>\beta_2 \Rightarrow \eigenspace(\alpha,\beta_1) \subseteq
\eigenspace(\alpha,\beta_2) \;.
$$
Moreover, for any sequence $\alpha_n,\beta_n \to 0$ we have
$$\eigenspace\setminus Sing =
\bigcup\limits_{n} \eigenspace(\alpha_n,\, \beta_n) \;.
$$

\begin{lemma}\label{lem:ler meas bnd}
For $f\in \eigenspace(\alpha,\, \beta)$ and
$0<\epsilon < \alpha$, we have
\begin{equation*}
\leray_\epsilon(f) <\frac{ d^{3/2}}{\beta}
2\sqrt{\eigmax}
\end{equation*}
where
$$\eigmax=\max\{|\lambda|^2:\lambda\in \Lambda\}\;.$$
\end{lemma}

We will first treat the one variable ($d=1)$ case and state it as
a separate lemma (cf \cite[Lemma 2]{Kac}):
\begin{lemma}\label{lem:Kac}
Let $g(t)$
be a trigonometric polynomial of degree at most $M$ so that there are
$\alpha>0$, $\beta>0$ such that
$|g'(t)|>\beta$ whenever $|g(t)|<\alpha$.
Then for all $0<\epsilon<\alpha$ we have
$$
\frac 1 {2\epsilon}\meas\{t: |g(t)|<\epsilon \} <\frac{2M}{\beta} \;.
$$
\end{lemma}
\begin{proof}
Decompose the open set $ \{t: |g(t)|<\epsilon \}$ as a disjoint union of
open intervals  $(a_k,b_k)$ (with $a_k<b_k$) and such that on each
such interval, $g'$
has constant sign, that is either $g'>\beta$ or $g'<-\beta$.
We will show that the length $b_k-a_k$ of
each such interval is at most $2\epsilon/\beta$ and that there are at
most  $2M$ such intervals.

Suppose that on  $(a_k,b_k)$,  $g'>\beta$; then $g$ is increasing,
and $g(a_k)=-\epsilon$, $g(b_k)=+\epsilon$. 
Then the length of the  interval is
\begin{equation*}
\begin{split}
b_k-a_k &= \int_{a_k}^{b_k} \frac {g'(t)}{g'(t)} dt
 < \frac {1}{\beta} \int_{a_k}^{b_k} g'(t)dt  \\
&= \frac{g(b_k)-g(a_k) }{\beta} =\frac {2\epsilon}{\beta}\;.
\end{split}
\end{equation*}

Likewise, if $g'<-\beta$ on $(a_k,b_k)$ then $g(a_k)=+\epsilon$,
$g(b_k)=-\epsilon$,  
and 
\begin{equation*}
\begin{split}
b_k-a_k &= \int_{a_k}^{b_k} \frac {-g'(t)}{-g'(t)} dt
< \frac 1\beta \int_{a_k}^{b_k} -g'(t)dt\\
&  = \frac{g(a_k)-g(b_k)}{\beta} =
\frac {2\epsilon}{\beta}
\end{split}
\end{equation*}
as required.

In both cases, each interval has  an endpoint where
$g(t)=+\epsilon$, and hence  the number
of such intervals is bounded by the number of solutions of
$g(t)=+\epsilon$ which is at most $2M$ since $g$ is a  trigonometric
polynomial of degree at most $M$.
\end{proof}

We  now prove Lemma~\ref{lem:ler meas bnd} by reduction to the case $d=1$:
\begin{proof}
Decompose the set $\{x: |f(x)|<\epsilon \}$ as a union $\cup_{j=1}^d W_j$
where
$$W_j = \{y: |f(y)|<\epsilon, \quad
|\frac{\partial f}{\partial x_j}(y)|\geq |\frac{\partial f}{\partial
x_k}(y)| \quad \forall k\neq j \}
$$
and it suffices to show that
$$
\meas(W_j) < 2\epsilon \frac{\sqrt{d}}{\beta} 4\sqrt{\eigmax}\;.
$$

For simplicity we fix $j=1$. On $W_1$, we have
$$ |\frac{\partial f}{\partial x_1}(y)| >\frac{\beta}{\sqrt{d}}$$
since $|f(y)|<\epsilon<\alpha$ implies
(recall $f\in \eigenspace(\alpha,\beta)$)
$$\beta^2<|\nabla f(y)|^2  = \sum_{k=1}^d
|\frac{\partial f}{\partial x_k}(y)|^2 \leq
d |\frac{\partial f}{\partial x_1}(y) |^2 \;.
$$

For $y\in \TT^{d-1}$ set
$$I(y) = \{t\in \TT^1: (t,y)\in W_1\}$$
which is a subset of $\TT^1$. Then slice-integration gives
$$
\meas(W_1) = \int_{\TT^{d-1}} \meas(I(y)) dy
$$
and so it suffices to show
\begin{equation*}
\meas(I(y)) < 2\epsilon \frac{\sqrt{d}}{\beta} 4 \sqrt{\eigmax} \;.
\end{equation*}

Now on $I(y)$, the one-variable trigonometric polynomial
$g(t):=f(t,y)$ satisfies $|g(t)|=|f(t,y)|<\epsilon$, and
$$
|g'(t)| = |\frac{\partial f}{\partial x_1}(t,y) | >
\frac{\beta}{\sqrt{d}} \;.
$$
Moreover $g(t)$ is of degree at most $\sqrt{\eigmax}$ because
$$
f(t,y) = \sum_{\lambda\in \Lambda } a_\lambda e(\lambda_1 t+
\sum_{j=2}^d \lambda_j y_j)
$$
and for all frequencies in the sum we have $\lambda_1^2 \leq
|\lambda|^2  \leq \eigmax$.
Thus by Lemma~\ref{lem:Kac} we find that
$\meas (I(y)) < 2\epsilon\frac{\sqrt{d}}{\beta} 2\sqrt{\eigmax}$
as required.
\end{proof}

\section{The expected value of $\leray$}
\label{sec:expect}

In this section, we give a formula for the expected value of
$\leray(f)$: 
\begin{theorem}\label{prop:exp nod leb} 
Suppose that $\Lambda$ is symmetric and
satisfies the nondegeneracy condition~\eqref{condition 1a}. 
Then the Leray measure $\leray (f)$
is integrable (with respect to the Gaussian measure),  and
\begin{equation}
\label{eq:exp nod meas leb}
\E( \leray ) =\frac{1}{\sqrt{2\pi}}.
\end{equation}
\end{theorem}

\subsection{A formal treatment}
To compute the expectation of $\leray(f)$, we formally write it as
$$\leray(f) = \int_{\TT^d} \delta(f(x))dx$$
and hence formally
$$\E(\leray(f)) = \E(\int_{\TT^d} \delta(f(x))dx) = \int_{\TT^d}
\E(\delta(f(x)) dx
$$
Now for each fixed $x\in \TT^d$, the random variable $f(x)$ is a sum
of Gaussians hence is itself a Gaussian whose mean is zero and
variance is computed to be unity. Hence the expected value
$\E(\delta(f(x))$ should be
$$\E(\delta(f(x)) = \int_{-\infty}^\infty
\delta(a)e^{-a^2/2}\frac{da}{\sqrt{2\pi}}  = \frac 1{\sqrt{2\pi}}
$$
which gives the result $\E(\leray) = 1/\sqrt{2\pi}$.
Justifying this simple manipulation in a rigorous fashion turns out to
be rather tedious will be done below, with some parts relegated to an
appendix.

\subsection{A rigorous proof}

The Leray measure $\leray(f)$ is defined outside of the singular set,
which has measure zero in $\eigenspace$, in fact forms a closed subset
of codimension  $\geq 1$ (Lemma~\ref{lem:sing codim 1}).
We compute the expectation of the nodal measure $\leray$
as follows:
We consider the increasing sequence of open subsets
$\eigenspace(\frac 1n, \frac 1n)$, $n=1,2,\dots$, whose union is the set of
nonsingular elements $\eigenspace\backslash Sing$.
We  choose subsets  $H_n \subset \eigenspace(\frac 1n, \frac 1n)$
which are (finite) unions of  {\em disjoint} open balls, so that
$$\bigcup_n H_n = \eigenspace \backslash Sing$$
and in fact the $H_n$ exhaust almost all nonsingular $f$'s, in the
sense that $\mu(H_n) \to 1$. (This is possible by Vitali's covering
theorem). 
We will show that  the limit
$$
\E(\leray) = \lim_n \int_{H_n} \leray(f)d\mu(f)
$$
exists and equals $1/\sqrt{2\pi}$.

By definition,
$$
\int_{H_n} \leray(f) d\mu(f) = \int_{H_n} \lim_{\epsilon\to 0}
\leray_\epsilon(f) d\mu(f) \
$$
where
$$
\leray_\epsilon(f):=
\frac 1{2\epsilon} \int_{\TT^d} \chi(\frac{f(x)}{\epsilon}) dx \;.
$$
By Lemma~\ref{lem:ler meas bnd}, on $H_n$,
$\leray_\epsilon(f) \leq c_n $
is  bounded uniformly for all $\epsilon<\frac 1n$.
Thus by the dominated convergence theorem we can exchange limits:
$$
\int_{H_n} \lim_{\epsilon\to 0} \leray_\epsilon(f) d\mu(f) =
\lim_{\epsilon\to 0} \int_{H_n} \leray_\epsilon(f) d\mu(f) \;.
$$
On the integral, we use Fubini's theorem to change the order of
integration
$$
 \int_{H_n} \leray_\epsilon(f) d\mu(f) =
\int_{\TT^d}\Bigg(
\frac 1{2\epsilon} \int_{H_n} \chi(\frac{f(x)}{\epsilon})
d\mu(f) \Bigg) dx \;.
$$

For the inner integral, we note that for each $x$, $f(x)$ is a
Gaussian random variable of mean zero and variance $\E(f(x)^2)=1$ and
hence setting
$$\Px^a=\{f\in\eigenspace: f(x)=a\}$$
which is an affine hyperplane of $\eigenspace$ of codimension one,
we have
$$
\frac 1{2\epsilon} \int_{H_n} \chi(\frac{f(x)}{\epsilon})  d\mu(f) =
\frac 1{2\epsilon}\int_{|a|<\epsilon} \mu_x^a(\Px^a \cap H_n)
e^{-a^2/2}\frac{da}{\sqrt{2\pi}}
$$
where $\mu_x^a$ is the induced Gaussian probability measure on the
hyperplane $\Px^a$.
Thus
$$
\int_{H_n} \leray(f)d\mu(f) = \lim_{\epsilon \to 0} \int_{\TT^d}
\frac 1{2\epsilon}\int_{|a|<\epsilon} \mu_x^a(\Px^a \cap H_n)
e^{-a^2/2}\frac{da}{\sqrt{2\pi}} dx \;.
$$


Now the function
$$
\mu_x^a(\Px^a \cap H_n)
$$
is bounded by $\mu_x^a(\Px^a) = 1$ and is {\em continuous} in both $a$
and in $x$ because we chose $H_n$ to be a
disjoint union of balls, and the volume of the
intersection of a hyperplane with this kind of nice set is a continuous
function of the hyperplane (since this is true for a ball).
Hence we may move the  limit  $\epsilon\to 0$ inside the integral over
$\TT^d$,  and find, by the
fundamental theorem of calculus, that
$$
\lim_{\epsilon \to 0} \frac 1{2\epsilon} \int_{|a|<\epsilon}
\mu_x^a(\Px^a \cap H_n) e^{-a^2/2}\frac{da}{\sqrt{2\pi}}
= \frac 1{\sqrt{2\pi}} \mu_x^0(\Px^0 \cap H_n) \;.
$$
Thus we find  that
$$\int_{H_n} \leray(f) d\mu(f) = \frac 1{\sqrt{2\pi}} \int_{\TT^d}
\mu_x^0(\Px^0 \cap H_n) dx\;.
$$

Now the functions
$$g_n(x):= \mu_x^0(\Px^0 \cap H_n)$$
are {\em continuous} in $x$, and are bounded: $g_n(x)\leq
\mu_x^0(\Px^0)=1$ 
and moreover for each $x$ their limit is
$$\lim_{n\to\infty} g_n(x) = \mu_x^0(\Px^0)=1$$
because by Proposition~\ref{lem:ssing zero meas Px} 
the singular set has measure 
zero in $\Px^0$ for each $x$ and the  $H_n$ exhaust  all the
nonsingular elements up to measure zero.

Thus we may in taking the limit $n\to \infty$ move the limit under the
integral to get
\begin{equation*}
\begin{split}
\lim_n \int_{H_n} \leray(f) d\mu(f) &=
\lim_n \frac 1{\sqrt{2\pi}}\int_{\TT^d}g_n(x) dx  \\
&= \frac 1{\sqrt{2\pi}} \int_{\TT^d} \lim_n g_n(x) dx =
\frac 1{\sqrt{2\pi}}
\end{split}
\end{equation*}
as required. \qed

\section{A formula for the variance of $\leray$}\label{sec:variance1}

In this section we give a formula for the variance of $\leray(f)$ in
terms of the two-point function 
$$
u(x,y) = \E(f(x) f(y))= \frac 1{\Ndim} \sum_{\lambda\in \Lambda}
\cos 2\pi \langle \lambda,z\rangle \;.
$$
The main result of this section is
\begin{theorem}\label{thm:variance formula}
Let $d\geq 2$. For any 
symmetric set of frequencies $\Lambda\subset \Z^d$ satisfying
the non-degeneracy condition~\eqref{condition 1a}, 
the second moment of $\leray$ is given by
$$
\E(\leray^2) = \frac 1{2\pi}\int_{\TT^d} \frac{dz}{\sqrt{1-u(z)^2}} \;.
$$
\end{theorem}

Thus the variance of $\leray$ is
$$
\var(\leray) =\frac 1{2\pi}\int_{\TT^d} \frac{dz}{\sqrt{1-u(z)^2}}
-\frac 1{2\pi} \;.
$$

\subsection{A formal derivation} \label{sec:formal}
It is simple to formally derive Theorem~\ref{thm:variance formula}:
Writing $\leray(f) =\int_{\TT^d} \delta(f(x))dx$ we have
\begin{equation*}
\begin{split}
\E(\leray^2) &=
\E\left(\int_{\TT^d}\int_{\TT^d} \delta(f(x))\delta(f(y)) dx dy\right)
\\&= \int_{\TT^d}\int_{\TT^d}
\E\Big(\delta(f(x))\delta(f(y))\Big) dxdy \;.
\end{split}
\end{equation*}
Now replace the vector $(f(x), f(y))$ by a  Gaussian vector
$a=(a_1,a_2)$ with
covariance matrix
\begin{equation}
\label{eq:Sigma}
\begin{pmatrix}
\E(f(x)^2)& \E(f(x)f(y)) \\ \E(f(y)f(x))& \E(f(y)^2)
\end{pmatrix}
= \begin{pmatrix}
1 &u(x-y) \\ u(y-x) &1
\end{pmatrix}
= \Sigma(x-y)
\end{equation}
whose determinant is $\det\Sigma(x-y)=1-u(x-y)^2$.
Thus
\begin{equation*}
\begin{split}
\E(\delta(f(x))\delta(f(y))) &= \iint_{\R^2} \delta(a_1)\delta(a_2)
\frac{e^{-\frac 12 a\Sigma^{-1} a^T}}{\sqrt{\det\Sigma}}
\frac{da_1da_2}{2\pi} \\
& = \frac 1{2\pi}\frac 1{\sqrt{1-u(x-y)^2}} \;.
\end{split}
\end{equation*}
This gives
$$\E(\leray^2) = \frac 1{2\pi} \int_{\TT^d}\int_{\TT^d} \frac
1{\sqrt{1-u(x-y)^2}} dxdy =  \frac 1{2\pi} \int_{\TT^d} \frac
1{\sqrt{1-u(z)^2}} dz
$$
as claimed.
The rigorous proof of this formula takes up the rest of the section.

\subsection{Integrability of the kernel}

\begin{lemma}\label{quadratic form}
Let $\Lambda \subset \mathbb R^d$ be invariant under permutations
and coordinate sign changes.   Then
$$\sum_{\lambda \in \Lambda} \langle \lambda ,\xi \rangle^2 = \frac 1d
\sum_{\lambda\in \Lambda} ||\lambda||^2 \cdot ||\xi||^2
$$
\end{lemma}
\begin{proof}
We  write the quadratic form in the LHS as
$$ Q(\xi) =\sum_{\lambda \in \Lambda} \langle \lambda ,\xi \rangle^2  =
\sum_{i,j=1}^d a_{ij} \xi_i \xi_j $$ 
where 
$$
a_{ij} =\sum_{\lambda\in \Lambda} \lambda_i\lambda_j \;.
$$


If $i\neq j$ use the symmetry under the sign change of the
$i$-th coordinate to change variables and deduce that $a_{ij}=0$.
For $i=j$ we find
$$a_{ii} = \sum_{\lambda\in \Lambda} \lambda_i^2$$
and the latter sum is independent of $i$ since $\Lambda$ is
symmetric under permutations; hence we may average the RHS over
$i$ to find
$$
a_{ii}=\frac 1d \sum_{i=1}^d \sum_{\lambda\in \Lambda} \lambda_i^2 =
\frac 1d \sum_{\lambda\in \Lambda} ||\lambda||^2
$$
as required.
\end{proof}

\begin{lemma}\label{lem:integrable}
For $d>1$, the kernel $1/\sqrt{\det \Sigma(z)}= \ 1/\sqrt{1-u(z)^2}$ is
integrable on $\TT^d$.
\end{lemma}
\begin{proof}
We need to check near the zeros of $\det \Sigma(z)$, that is at points
where $u(z)=\pm 1$. By Lemma~\ref{lem:u(x)!=u(y) a.a}  this 
implies that $z$ lies in the finite set $\B/\Z^d$. 
At such points $z_0$, all the 
cosines $\cos2\pi \langle \lambda, z_0 \rangle$ have the same value,
which is either $+1$ or $-1$, and expanding in a small neighbourhood
we have
$$
\cos 2\pi \langle \lambda, z \rangle
\sim \pm(1 - \frac 12 \langle\lambda, z-z_0 \rangle^2 ) \;.
$$
Thus
$$
\det \Sigma(z) = 1-u(z)^2
\sim  \frac 1{2\Ndim} \sum_{\lambda\in\Lambda}
\langle\lambda, z-z_0 \rangle^2 \;.
$$
By Lemma~\ref{quadratic form}, we thus have
$$
\det\Sigma(z) \sim
\frac 1{2d} \left(\frac 1{\Ndim}\sum_{\lambda\in \Lambda} |\lambda|^2 \right)
|z-z_0|^2
$$
and therefore
$$
\frac 1{\sqrt{\det\Sigma(z)}} \sim const. \frac 1{|z-z_0|}
$$
near $z_0$, which is integrable if (and only if) $d>1$.
\end{proof}

\subsection{Proof of Theorem~\ref{thm:variance formula}}
We have
$$
\int_{H_n} \leray(f)^2 d\mu(f) = \int_{H_n}
\lim_{\epsilon_1,\epsilon_2 \to 0} \leray_{\epsilon_1}(f)
\leray_{\epsilon_2}(f) d\mu(f) \;.
$$
By Lemma~\ref{lem:ler meas bnd}  and the dominated convergence
theorem, we may take the
limit outside the integral sign and get
$$
\int_{H_n} \leray(f)^2 d\mu(f) = \lim_{\epsilon_1,\epsilon_2 \to 0}
\int_{H_n} \frac 1{4\epsilon_1 \epsilon_2} \iint_{\TT^d\times \TT^d}
\chi(\frac{f(x)}{\epsilon_1}) \chi(\frac{f(y)}{\epsilon_2}) dx dy
d\mu(f)
$$
which by Fubini's theorem and the change of variable $y=x+z$, equals
$$
\lim_{\epsilon_1,\epsilon_2 \to 0}
\iint_{\TT^d\times \TT^d} \left( \frac 1{4\epsilon_1 \epsilon_2}
\int_{H_n}  \chi(\frac{f(x)}{\epsilon_1})
\chi(\frac{f(x+z)}{\epsilon_2})
d\mu(f)
\right) dx dz \;.
$$

\subsubsection{Excising the singular set}
Fix $\epsilon_1,\epsilon_2>0$ and let
$S(\epsilon_1,\epsilon_2)\subset \TT^d$ be a subset
of measure at most $(\epsilon_1 \epsilon_2)^2$ surrounding
the finitely many points of $\B/\Z^d$.
Then using $\chi\leq 1$ we have
\begin{multline*}
\int_{\TT^d} \int_{S(\epsilon_1, \epsilon_2)}
\left( \frac 1{4\epsilon_1 \epsilon_2}
\int_{H_n}  \chi(\frac{f(x)}{\epsilon_1})
\chi(\frac{f(x+z)}{\epsilon_2})
d\mu(f)
\right) dx dz \\
\leq  \frac {\meas S(\epsilon_1,\epsilon_2)}{4\epsilon_1 \epsilon_2}
<\epsilon_1 \epsilon_2
\end{multline*}
and hence the in the limit $\epsilon_1,\epsilon_2\to 0$ this gives zero
contribution. Thus
\begin{multline*}
\int_{H_n} \leray(f)^2d\mu(f) \\
= \lim_{\epsilon_1,\epsilon_2\to 0}
\int_{\TT^d}\int_{\TT^d\backslash S(\epsilon_1,\epsilon_2)}
\left( \frac 1{4\epsilon_1 \epsilon_2} \int_{H_n}
\chi(\frac{f(x)}{\epsilon_1})\chi(\frac{f(x+z)}{\epsilon_2}) d\mu(f)
\right)dxdy \;.
\end{multline*}

\subsubsection{Gaussian integration}
For fixed $\epsilon_1,\epsilon_2>0$ we evaluate the inner integral
as in the formal derivation of \S~\ref{sec:formal} by replacing the vector
$(f(x),f(y))$ by a Gaussian vector $(a_1,a_2)\in \R^2$ with covariance matrix
$\Sigma(z)$ given in \eqref{eq:Sigma}. For $x-y\notin \B/\Z^d$ 
and $a=(a_1,a_2)\in \R^2$, set
$$
\Pxy^a = \{f\in \eigenspace: f(x)=a_1, f(y)=a_2 \} \;,
$$
which is an affine subspace of  codimension two.
Let $\mu_{x,y}^a$ be
the induced Gaussian probability measure on $\Pxy^a$.
Then for $z=x-y\notin \B/\Z^d$,
\begin{multline*}
  \frac 1{4\epsilon_1 \epsilon_2} \int_{H_n}
\chi(\frac{f(x)}{\epsilon_1})\chi(\frac{f(x+z)}{\epsilon_2}) d\mu(f)=\\
\frac 1{ \sqrt{\det \Sigma(z)}}
 \frac 1{4\epsilon_1 \epsilon_2}
\iint_{|a_1|<\epsilon_1, |a_2|<\epsilon_2}
e^{-\frac 12 a\Sigma^{-1}(z) a^T} \mu_{x,x+z}^a(\Pxz^a \cap H_n)
\frac{ da_1 da_2}{2\pi} \;.
\end{multline*}
Thus we find
$$
\int_{H_n} \leray(f)^2 d\mu(f) = \lim_{\epsilon_1,\epsilon_2\to 0}
\int_{\TT^d}\int_{\TT^d\backslash S(\epsilon_1,\epsilon_2)}
K_n(x,x+z;\epsilon_1,\epsilon_2) dxdz
$$
where
\begin{multline*}
K_n(x,x+z;\epsilon_1,\epsilon_2)=\\
 \frac 1{ \sqrt{\det \Sigma(z)}}
 \frac 1{4\epsilon_1 \epsilon_2}
\iint_{|a_1|<\epsilon_1, |a_2|<\epsilon_2}
e^{-\frac 12 a\Sigma^{-1}(z) a^T} \mu_{x,x+z}^a(\Pxz^a \cap H_n)
\frac{ da_1 da_2}{2\pi} \;.
\end{multline*}

\subsubsection{Excising more points}
Fix $\delta>0$, and for $\epsilon_1, \epsilon_2$ sufficiently small
fix a set $D\subset \TT^d$ so that
\begin{enumerate}
\item $S(\epsilon_1,\epsilon_2) \subset D$
\item $D$ contains the measure zero set of $z=x-y$ for which
Proposition~\ref{lem:SSing in Pxy is zero} fails to hold.
\item $\int_{D} \frac{dz}{\sqrt{\det \Sigma(z)}} <\delta$.
\end{enumerate}
Then we can bound
$$
K_n(x,x+z;\epsilon_1,\epsilon_2)\leq \frac 1{2\pi
\sqrt{\det \Sigma(z)}}
$$
by using $ e^{-\frac 12 a\Sigma^{-1}(z) a^T}\leq 1$ and
$ \mu_{x,x+z}^a(\Pxz^a \cap H_n) \leq  \mu_{x,x+z}^a(\Pxz^a )
=1$.
Thus
$$
\int_{H_n} \leray(f)^2 d\mu(f) =
\lim_{\epsilon_1,\epsilon_2\to 0}
\int_{\TT^d}\int_{\TT^d\backslash D} K_n(x,x+z;\epsilon_1,\epsilon_2) dxdz
 + O(\delta) \;,
$$
with the implied constant in $O(\delta)$  independent of $n$.

\subsubsection{A switch of limit and integration}
Since $K_n$ is dominated by $1/\sqrt{\det \Sigma(z)}$, which is
integrable by Lemma~\ref{lem:integrable}, we may use the dominated convergence
theorem to switch the limit $\epsilon_1,\epsilon_2\to 0$ and the
integral to get
$$
\int_{H_n} \leray(f)^2 d\mu(f) =
\int_{\TT^d}\int_{\TT^d\backslash D} \lim_{\epsilon_1,\epsilon_2\to 0}\;
K_n(x,x+z;\epsilon_1,\epsilon_2) \; dxdz  + O(\delta)\;.
$$
where the implied constant is independent of $n$.

\subsubsection{Taking the limit $\epsilon_1,\epsilon_2 \to 0$}
The  function
$$
(x,z,a)\mapsto e^{-\frac 12 a\Sigma^{-1}(z) a^T}
\mu_{x,x+z}^a(\Pxz^a \cap H_n)
$$
is continuous 
on $\TT^d\times \TT^d\backslash D\times \R^2$
by construction of $H_n$ to have continuous intersection with
hyperplanes of fixed dimension.
Thus for $z\notin D$, we may use the fundamental theorem of calculus
to get
\begin{multline*}
\lim_{\epsilon_1,\epsilon_2\to 0}
 \frac 1{4\epsilon_1 \epsilon_2}
\iint_{|a_1|<\epsilon_1, |a_2|<\epsilon_2}
e^{-\frac 12 a\Sigma^{-1}(z) a^T} \mu_{x,x+z}^a(\Pxz^a \cap H_n)
da_1 da_2 \\
= \mu_{x,x+z}^0 (\Pxz^0\cap H_n) \;.
\end{multline*}
Therefore for $z\notin D$,
$$
\lim_{\epsilon_1,\epsilon_2\to 0} K_n(x,x+z;\epsilon_1,\epsilon_2) =
 \frac{\mu_{x,x+z}^0 (\Pxz^0\cap H_n)}
{2\pi \sqrt{\det\Sigma(z)}}  \;.
$$
This gives
$$
\int_{H_n} \leray(f)^2 d\mu(f) =
\int_{\TT^d}\int_{\TT^d\backslash D}\frac{\mu_{x,x+z}^0 (\Pxz^0\cap H_n)}
{2\pi \sqrt{\det\Sigma(z)}}  dx dz +O(\delta)\;.
$$

\subsubsection{The limit $n\to \infty$}
Taking now the limit $n\to \infty$, and using continuity of
$\mu_{x,x+z}^0 (\Pxz^0\cap H_n)$
on $\TT^d\times \TT^d\backslash D$ (which is is due to
the construction of $H_n$) and using
Proposition~\ref{lem:SSing in Pxy is zero}   to
guarantee that for $z\notin D$, the intersection of $\Pxz^0$ with the
singular set has measure zero in $\Pxz^0$, we find
$$\lim_{n\to \infty} \mu_{x,x+z}^0 (\Pxz^0\cap H_n) =
\mu_{x,x+z}^0 (\Pxz^0)=1
$$
and thus
$$
\lim_{n\to \infty} \int_{H_n} \leray(f)^2 d\mu(f) =
\int_{\TT^d\backslash D} \frac {dz}{2\pi \sqrt{\det \Sigma(z)}}
+O(\delta) \;.
$$
Since $\delta>0$ is arbitrary and $1/\sqrt{\det\Sigma(z)}$ is
integrable on $\TT^d$, we finally conclude that
$$
\E(\leray^2) = \lim_{n\to \infty} \int_{H_n} \leray(f)^2 d\mu(f) =
\frac 1{2\pi}\int_{\TT^d} \frac {dz}{ \sqrt{\det \Sigma(z)}} \;.
$$
This concludes the proof of Theorem~\ref{thm:variance formula}.
\qed

\section{The asymptotics of the variance} \label{sec:asymp}

In the previous section we showed that the second moment of the Leray
measure for the ensemble of trigonometric polynomials associated to any
symmetric set of frequencies is given by
\begin{equation}\label{eq:second mom}
 \E(\leray^2) = \frac 1{2\pi} \int\limits_{\T^d}
\frac{dx}{\sqrt{1-u^2(x)}}
\end{equation}
where $u(x) = \frac{1}{\Ndim} \sum\limits_{\lambda\in\Lambda}
\cos 2\pi \langle \lambda,\, x\rangle$ is the two-point
function of the process.

 From now on,  we specialize to the case that
$$
\Lambda=\{\lambda\in \Z^d: |\lambda|^2=\eigenvalue \}\;.
$$
In this section we show:
\begin{proposition} \label{prop I(Lambda)}
The second moment of $\leray(f)$ is given by
\begin{equation*}
\E(\leray^2) = \frac 1{2\pi}+\frac 1{4\pi\Ndim} +
O\bigg(\int_{\T^d} u(x)^4 dx\bigg) \;.
\end{equation*}
\end{proposition}

In section \S~\ref{sec:fourth moment} we will see that for $d=2$ and
$d\geq 5$, the fourth moment of $u$ is negligible relative to
$1/\Ndim$ and hence we will obtain
$$\var(\leray) \sim \frac 1{4\pi \Ndim}$$
as $\Ndim\to \infty$, which is Theorem~\ref{thm:var}.

%
%

We now set about the proof of Proposition~\ref{prop I(Lambda)}.

\subsection{Singular points}
\begin{defn} A point $x\in \T^d$ is a positive singular point if
  there is a set of frequencies $\Lambda_x\subset \Lambda$ with
  density $\frac{|\Lambda_x|}{|\Lambda|}>1-\frac 1{4d}$ for which
$\cos 2\pi\langle \lambda, x \rangle >3/4$ for all $\lambda\in \Lambda_x$.
Similarly we define a negative singular point to be a point $x$ where
  there is a set
$\tilde\Lambda_x\subset\Lambda$   of density $>1-\frac 1{4d}$ for which
$\cos 2\pi\langle \lambda, x\rangle <-3/4$ for all 
$\lambda\in \tilde\Lambda_x$.
\end{defn}
An example is the origin, where $\cos 2\pi \langle\lambda, 0\rangle=1$.

Let $M\approx \sqrt{\eigenvalue}$ be a large integer\footnote{It
suffices to take $M=\lfloor 16\pi \sqrt{d}\sqrt{E} \rfloor$.}.
We decompose the unit cube (the torus) as a disjoint union (with
boundary overlaps) of $M^d$ closed cubes $I_{\vec k}$ of side
length $1/M$ centered at $\vec k/M$, $\vec k\in \Z^d$.

\begin{defn}A cube $I_{\vec k}$ is a positive (resp. negative)
singular cube if it contains a positive (resp. negative) singular
point.
\end{defn}

\begin{lemma}\label{cos large in cube}
For a positive (respectively, negative) singular cube $I$, there is a subset of frequencies
$\Lambda_I\subset \Lambda$ with  with
  density $\frac{|\Lambda_I|}{|\Lambda|}>1-\frac 1{4d}$ for which
$\cos 2\pi\langle \lambda, y\rangle >1/2$ (respectively, $\cos
2\pi\langle \lambda, y\rangle <-1/2$) for all $y\in I$ and all
$\lambda\in \Lambda_I$.  
\end{lemma}
\begin{proof}
Let $x\in \Lambda$ be a positive singular point, and let
$\Lambda_I=\Lambda_x$ be the set of frequencies for which
$\cos 2\pi\langle \lambda, x\rangle >3/4$. It suffices to show that if
$|y-x| \ll 1/M$ then $\cos 2\pi\langle \lambda, y\rangle >1/2$ for all
$\lambda\in \Lambda_x$.  

By the mean value theorem and Cauchy-Schwartz,
\begin{equation*}
\begin{split}
|\cos 2\pi \langle\lambda, y\rangle-\cos 2\pi\langle\lambda, x\rangle|
&= | \langle -2\pi\sin 2\pi\langle\lambda, \xi\rangle 
\lambda,x-y \rangle| \\
&\leq 2\pi |\lambda| |x-y| \ll \frac{\sqrt{\eigenvalue}}{M}
\end{split}
\end{equation*}
and hence if $M\gg \sqrt{\eigenvalue}$ (all implied constants are
absolute, depending only on the dimension $d$) and $\cos 2\pi
\langle \lambda, x\rangle >3/4$ then
$$
\cos 2\pi \langle \lambda, y\rangle \geq \cos 2\pi \langle \lambda,
x\rangle- |\cos 2\pi \langle\lambda, y\rangle-\cos 2\pi \langle
\lambda, x\rangle |
> \frac 34 - \frac 14 = \frac 12
$$
as required. The case of negative singular cubes is analogous.
\end{proof}
As Lemma~\ref{cos large in cube} shows, singular cubes cannot be
both positive and negative.

Let  $B$ be the union of all singular cubes. Since the volume of each
cube is $1/M^d$, the number of such cubes is  $M^d\meas(B)$.

\begin{lemma}\label{lem:bounding u}
i) If $x\notin B$ then $|u(x)|<1-\frac 1{16d}$. 

ii) If $x\in B$ then $|u(x)|>\frac 12-\frac 3{8d}\geq \frac 1{16}$.

iii) $\meas(B) \leq 16^4 \int_{\TT^d} u(x)^4dx$. 
\end{lemma}
\begin{proof}
i) If $x\notin B$, then $x$ is neither a positive nor
a negative singular point, hence 
there are subsets $\Lambda',\Lambda''\subset
\Lambda$ each of density $> \frac 1{4d}$ for which $\cos 2\pi
\langle \lambda, x\rangle\leq \frac 34$ for all $\lambda\in \Lambda'$  and
$\cos 2\pi \langle\lambda, x\rangle\geq -\frac 34$ for all $\lambda\in
\Lambda''$. Hence
\begin{equation*}\begin{split}
u(x) &= \frac 1\Ndim\sum_{\lambda\in \Lambda'}\cos 2\pi\langle
\lambda, x\rangle + 
\frac 1\Ndim\sum_{\lambda\notin \Lambda'}\cos 2\pi\langle
\lambda, x\rangle  \\
&\leq
\frac 34\frac {|\Lambda'|}{\Ndim} + \frac{\Ndim-|\Lambda'|}{\Ndim} \\
&=  1-\frac 14\frac {|\Lambda'|}{\Ndim} < 1 -\frac 1{16d} \;.
\end{split}\end{equation*}
Likewise, using $\Lambda''$ instead of $\Lambda'$, we also have
$u(x)> - 1 +\frac 1{16d}$ and hence $|u(x)|< 1 -\frac 1{16d}$.

ii) Suppose $x\in B$ lies in a positive singular cube. Then by
Lemma~\ref{cos large in cube} there is
$\Lambda'\subset \Lambda$, with 
$\frac {|\Lambda'|}{|\Lambda|}>1-\frac 1{4d}$, such that $\cos 2\pi
\langle \lambda,x \rangle >\frac 12$ for all $\lambda\in \Lambda'$. 
Hence 
\begin{equation*}
\begin{split}
u(x) &= \frac 1{\Ndim}\sum_{\lambda\in \Lambda'} \cos2\pi \langle
\lambda,x \rangle  + 
\frac 1{\Ndim}\sum_{\lambda\notin \Lambda'} \cos2\pi \langle
\lambda,x \rangle \\
& > \frac 1{\Ndim} \sum_{\lambda\in \Lambda'} \frac 12 +
\frac 1{\Ndim}\sum_{\lambda\notin \Lambda'} (-1) \\
&= \frac 12 \frac{|\Lambda'|}{\Ndim} -\frac{\Ndim -|\Lambda'|}{\Ndim}
=\frac 32 \frac{|\Lambda'|}{\Ndim} -1 \\
&>\frac 32(1-\frac 1{4d})-1 = \frac 12-\frac 3{8d} \geq \frac 1{16} \;.
\end{split}
\end{equation*}
Thus $u(x)>\frac 12-\frac 3{8d} \geq \frac 1{16} $. Likewise if $x$
lies in a negative singular cube we will find that $u(x)<-\frac
1{16}$ and hence for all $x\in B$ we have $|u(x)|>\frac 1{16}$. 

iii) follows from (ii) by a Chebyshev type inequality. 
\end{proof}

We separately compute the contributions  $I_B$, $I_{B^c}$, of the
singular set $B$ and its complement $B^c$ to \eqref{eq:second mom}.

\subsection{The contribution of $B^c$}
This will be the main term. For $x\notin B$, since $|u(x)|$ is bounded
away from $1$, we may use the Taylor expansion
$$\frac 1{\sqrt{1-u(x)^2}} = 1+\frac 12 u(x)^2 +O(u(x)^4)$$
(the implied constant independent of $\Lambda$!) to find
\begin{equation*}\begin{split}
I_{B^c} &=\frac 1{2\pi}  \int_{B^c} \frac {dx}{\sqrt{1-u(x)^2}}  =
\frac 1{2\pi} \int_{B^c} \left( 1+\frac 12 u(x)^2+O(u(x)^4) \right) dx\\
&=\frac 1{2\pi}  + \frac 1{4\pi}\int_{\T^d} u(x)^2 dx+
O(\meas(B)) +O(\int_{\T^d}
u(x)^4 dx)
\end{split}\end{equation*}
on using $|u(x)|\leq 1$; then since $\int_{\T^d}u(x)^2dx=\frac 1\Ndim$
and $ \meas(B)\ll \int_{\T^d} u^4$ by 
Lemma~\ref{lem:bounding u}(iii), we find
\begin{equation}\label{cont nonsing} 
I_{B^c} = \frac 1{2\pi} +\frac 1{4\pi\Ndim} + O\bigg(\int_{\T^d}
u^4\bigg) \;.  
\end{equation}

\subsection{The contribution of the singular set  $B$}
\label{subsec:contr sing}
To estimate $I_B$, we will show  that each integral over a single
singular cube contributes $O(1/M^{d-1}\sqrt{\eigenvalue})$. Since 
the number of singular cubes is $M^d \meas(B)$,  we will
find that the total contribution of $I_B$ is bounded by
\begin{equation}\label{cont sing cube}
I_B\ll \meas(B)\frac{M}{\sqrt{\eigenvalue}}\approx
\meas(B) \ll \int_{\T^d} u^4
\end{equation}
because we assume that $M\approx \sqrt{\eigenvalue}$. 
Together with \eqref{cont nonsing}, this will prove
Proposition~\ref{prop I(Lambda)}.  

\subsection{A bound for the Hessian of $u$ on a cube}
The Hessian  of $u$ is $H=(\frac{\partial^2 u}{\partial
x_i\partial x_j})$. We will need to know:
\begin{lemma} \label{Hessian}
The Hessian of $u$ at any point in a positive  singular cube is
negative definite and satisfies
$$
\xi^T H\xi\leq -\frac{\pi^2\eigenvalue}{2d}  ||\xi||^2 \;.
$$
Likewise for a negative singular cube the Hessian is positive
definite and satisfies $\xi^T H\xi\geq \frac{\pi^2 \eigenvalue}{2d}||\xi||^2$.
\end{lemma}
\begin{proof}
The Hessian $H_\lambda$ of $\cos 2\pi\langle \lambda, x\rangle$ is
given by 
$$
(H_\lambda)_{i,j}= -(2\pi)^2 \cos 2\pi \langle\lambda,x\rangle 
\lambda_i \lambda_j = -(2\pi)^2\cos 2\pi \langle\lambda, x\rangle( \lambda
\lambda^T)_{i,j}
$$
(if we think of $\lambda$ as a column vector) for which
$$ \xi^T H_\lambda \xi = - \cos 2\pi \langle\lambda, x \rangle
\langle \lambda, 2\pi\xi\rangle^2 \;.
$$
Let $\Lambda'\subset \Lambda$ be a set of frequencies of density
$> 1-\frac 1{4d}$ so that for all $x$ in the singular cube, and
all $\lambda\in \Lambda'$, we have $\cos 2\pi\langle \lambda, x\rangle>1/2$.
Then for $\lambda\in \Lambda'$ (the weak inequality is introduced
to cover the case that $\langle \lambda,\xi\rangle=0$)
$$ 
\xi^T H_\lambda \xi\leq -\frac 12 \langle\lambda,2\pi \xi\rangle^2 \;.
$$

For the remaining $\lambda\notin \Lambda'$, we use
$-\cos 2\pi\langle\lambda, x\rangle\leq 1$ to get 
$\xi^T H_\lambda\xi\leq \langle \lambda,2\pi \xi \rangle^2$. 
Hence the Hessian $H$ of $u$ at $x$ 
satisfies
\begin{equation*}
\begin{split}
\xi^T H \xi  &=\frac 1{\Ndim} \sum_{\lambda\in \Lambda} \xi^T
H_\lambda \xi \\
& \leq -\frac 12 \frac 1\Ndim\sum_{\lambda\in \Lambda'}
\langle   \lambda,2\pi\xi\rangle^2
+ \frac 1\Ndim\sum_{\lambda\notin \Lambda'} \langle
\lambda,2\pi\xi\rangle^2
\\
& = -\frac 12 \frac 1\Ndim\sum_{\mbox{ all } \lambda} \langle
\lambda,2\pi\xi \rangle^2 + \frac 32 \frac 1\Ndim \sum_{\lambda \notin
\Lambda'}\langle \lambda,2\pi\xi \rangle^2
\end{split}
\end{equation*}
for all $\xi$.
By Lemma~\ref{quadratic form} we have
$$
 -\frac 12 \frac 1\Ndim\sum_{\mbox{ all } \lambda}
\langle \lambda,2\pi\xi \rangle^2  = -\frac{2\pi^2\eigenvalue}{d}
||\xi||^2 \;.
$$
For the sum over $\lambda\notin \Lambda'$, use Cauchy-Schwartz to
write
$$
\langle \lambda,2\pi \xi\rangle^2 \leq 
4\pi^2\eigenvalue||\xi||^2
$$
and the sum over these $\lambda \notin\Lambda'$ is hence bounded
by
$$
\frac 32\frac {\Ndim-|\Lambda'|}\Ndim 4\pi^2\eigenvalue ||\xi||^2
\leq \frac {3\pi^2}{2d}\eigenvalue ||\xi||^2
$$
(since $\frac{|\Lambda'|}{\Ndim}\geq 1-\frac 1{4d}$).
Thus we find
$$
\xi^T H\xi \leq -\frac{\pi^2}{2d}\eigenvalue  ||\xi||^2
$$
as required.
\end{proof}

\subsection{The contribution of a singular cube}
To find the contribution to the integral of each singular cube
$I_k$, assume the cube contains a positive singular point.  

Pick a point $x_0\in I_k$ for which $u(x_0)$ is maximal in $I_k$.
Now use the Taylor expansion around $x_0$ with remainder
$$u(x) = u(x_0)+ \nabla u(x_0)\cdot (x-x_0) + R_2(x)
$$
where the remainder $R_2(x)$ can be given in terms of the Hessian
$H$
of $u$ as
$$ R_2(x) = \frac 12 (x-x_0)^T H(z)(x-x_0)$$
where $z$ is some point on the line segment between $x_0$ and $x$.
Since the cube is convex, $z$ also belongs to the singular cube.
Thus by Lemma~\ref{Hessian}, 
we have
$$R_2(x) \leq -\frac {\pi^2\eigenvalue}{4d} ||x-x_0||^2 \;.$$

The directional derivative at $x_0$ of $u$ in the direction of any
other point in the cube is nonpositive (since the function is
decreasing as we go from $x_0$ to nearby points in the cube) and
hence
$$\nabla u(x_0)\cdot (x-x_0)\leq 0$$
for all points $x$ in the cube, as this quantity is a positive
multiple of the directional derivative of $u$ at $x_0$ in the
direction of the line joining $x_0$ to $x$. Thus
\begin{equation*}\begin{split}
u(x) & = u(x_0) + \nabla u(x_0)\cdot (x-x_0) + \frac 12 (x-x_0)^T
H(z) (x-x_0)  \\
&\leq 1 +0 -\frac{\pi^2\eigenvalue}{4d}||x-x_0||^2 \;.
\end{split}\end{equation*}
Therefore 
$$ 1-u^2\gg \eigenvalue ||x-x_0||^2  $$
amd hence the integral over a positive singular cube is bounded by
$$
\int_{||x-x_0||\ll 1/M} \frac{dx}{\sqrt{\eigenvalue ||x-x_0||^2}} \ll \frac
1{\sqrt{\eigenvalue}} \int_0^{1/M} \frac{r^{d-1}dr}r \approx \frac
1{\sqrt{\eigenvalue}M^{d-1}} \;.
$$

The case of a negative singular cube is analogous; instead
of using a maximum of $u$ in the cube we take $x_0$ to be a
minimum of $u$ in the cube and show that $u(x)\geq -1+
\frac{\pi^2\eigenvalue}{4d}||x-x_0||^2$. 

Thus we have proved \eqref{cont sing cube} and hence are done with 
the proof of Proposition~\ref{prop I(Lambda)}. 

\section{Bounding the fourth moment of the two-point function}
\label{sec:fourth moment}
In this section we bound the fourth moment of the two-point function
$$
u(x)=\frac 1\Ndim\sum_{\lambda\in \Lambda}
e^{2\pi i\langle\lambda, x\rangle} \;.
$$
Note that
\begin{equation*}
\int_{\TT^d}u(x)^4dx = \frac
1{\Ndim^4}\#\{\lambda_1,\lambda_2,\lambda_3,\lambda_4 :
\lambda_1+\lambda_2=\lambda_3+\lambda_4 \} \;.
\end{equation*}
The number of solutions of the equation
\begin{equation}\label{eq: equation for u4}
\lambda_1+\lambda_2=\lambda_3+\lambda_4, \qquad \lambda_i\in \Lambda
\end{equation}
is at most $\Ndim^3$ since fixing three of the variables determines the
fourth one. Thus
\begin{equation}\label{general bound on m4}
\int u^4 dx\leq \frac 1\Ndim \;.
\end{equation}

This bound used no special property of the set of frequencies
$\Lambda$.
For the set $\Lambda_E=\{\lambda: |\lambda|^2=\eigenvalue\}$
we can do much better.


\begin{proposition}\label{prop:fourth moment}
i) In dimension $d=2$, we have
$$\int u^4 
\ll \frac 1{\Ndim^2} \;.
$$

ii) In dimension $d\geq 3$,
$$
\int_{\TT^d} u(x)^4dx \ll_\epsilon
\frac{\eigenvalue^{\frac{d-3}2+\epsilon}}{\Ndim^2}
$$
for all $\epsilon>0$.
\end{proposition}
To prove the proposition, we need to bound the number of solutions of
\eqref{eq: equation for u4}.
A simple geometric  argument pointed out by Zygmund \cite{Zygmund}
shows that  in dimension $d=2$, the only solutions of
\eqref{eq: equation for u4} are ``diagonal'' solutions, that is
$\lambda_1=\lambda_3$, or $\lambda_1+\lambda_2=0=\lambda_3+\lambda_4$
etcetera. This gives the required bound in two dimensions.

For higher dimensions,
we want to show that the number of solutions of
\eqref{eq: equation for u4} is
$\ll \Ndim^2\eigenvalue^{\frac{d-3}2+\epsilon}$.
Fix $\lambda_3,\lambda_4$. If $\lambda_3+\lambda_4=0$ then
$\lambda_1+\lambda_2=0$ and there are $\Ndim^2$ such pairs. So we may
ignore them and assume that $\nu:=\lambda_3+\lambda_4\neq 0$ and then we
wish to show that there are at most
$\eigenvalue^{\frac{d-3}2+\epsilon}$ choices of
of $\lambda_1,\lambda_2$ with $\lambda_1+\lambda_2=\nu$ given. Since
$\lambda_2=\nu-\lambda_1$ is determined by $\lambda_1$, we thus need
to show:

\begin{lemma}
Let $d\geq 3$ and $0\neq \nu\in \Z^d$.
Then the number of $\lambda\in \Z^d$ with
\begin{equation}\label{eq sim}
|\lambda|^2=\eigenvalue=|\nu-\lambda|^2
\end{equation}
is at most $c(\epsilon)\eigenvalue^{\frac{d-3}2+\epsilon}$ for all
$\epsilon>0$ with $c(\epsilon)>0$ independent of $\nu$.
\end{lemma}
\begin{proof}
To see this, rewrite the equations as
$$|\lambda|^2=\eigenvalue, \quad 2\langle \lambda,\nu\rangle = |\nu|^2
$$
or
$$
\sum_{j=1}^d x_j^2 = \eigenvalue,\quad 2\sum_{j=1}^d\nu_j x_j =
|\nu|^2 \;.
$$
Fix the last $d-3$ coordinates $x_4,\dots, x_d$ (there are at most
$\eigenvalue^{\frac{d-3}2}$ such choices) and lets count the number of
solutions of the resulting system of equations
\begin{equation}\label{three vars}
x_1^2+x_2^2+x_3^2=R,\quad \nu_1x_1+\nu_2x_2+\nu_3 x_3=S
\end{equation}
where  $R\leq \eigenvalue$ and $|\nu_i|,|S|\ll \eigenvalue$.
The number of solutions of \eqref{eq sim} is thus bounded by 
$\eigenvalue^{\frac{d-3}2}$ times the number of solutions of equations 
such as  \eqref{three vars}. So it suffices to show that the number of
solutions of  \eqref{three vars} is at most
$c(\epsilon)\eigenvalue^\epsilon$ uniformly in $\nu$.  

Solving  the linear equation for $x_3$ and substituting in the
quadratic equation gives an inhomogeneous quadratic equation
$$
ax_1^2+bx_1x_2 +c x_2^2 +dx_1 +ex_2 +f=0
$$
where all coefficients are integers which are at most polynomial in
$\eigenvalue$ and the homogeneous quadratic part is positive definite.
Then one may complete the square and change variables to get an
equation
$$x^2+Dy^2 = k$$
where $D>0$, and $D,k$ are polynomial in $\eigenvalue$. Thus the
number of solutions of  \eqref{three vars} is bounded by
the number $r_D(k)$ of representations of an integer
$k$ by the quadratic form $x^2+Dy^2$.

Now we claim that $r_D(k)$  is at most
\begin{equation}\label{estimate on r_D(k)}
r_D(k) \leq 6 \tau(k)
\end{equation}
where $\tau(k)$ is the number of divisors of $k$. Since
$\tau(k)\ll k^\epsilon$, $\forall \epsilon>0$,
this will imply that the number of solutions to \eqref{three vars} is at most
$c(\epsilon)\eigenvalue^\epsilon$ uniformly in $\nu$ and conclude the
proof of the lemma.

The uniform estimate \eqref{estimate on r_D(k)}
follows from factorization into prime ideals in the ring of integers of the
imaginary quadratic extension $\Q(\sqrt{-D})$:
Indeed, $r_D(k)$ is at most the number $\rho(k)$ of ideals
of norm $k$, times the number of units of the field, which is at most $6$.
Now the Dirichlet series $\zeta_D(s):=\sum_{k\geq 1}\rho(k)/k^s$  is
the Dedekind  zeta function of the field $\Q(\sqrt{-D})$, and by
class-field theory 
there is a factorization $\zeta_D(s) = \zeta(s)L(s,\chi)$ where
$\zeta(s)$ is the Riemann zeta function, and $L(s,\chi)$ is the
Dirichlet L-function associated to the quadratic character $\chi$
attached to $\Q(\sqrt{-D})$. Thus $\rho(k) = \sum_{m\mid k} \chi(m)$
and therefore $\rho(k)$ is bounded by the number $\tau(k)$ of divisors
of $k$. Thus $r_D(k)\leq 6 \tau(k)$.
\end{proof}



\vskip .5cm
\noindent{\bf Remark.}
For higher dimensions, one can improve 
on  the trivial bound \eqref{general bound on m4} by noting that
$u(x)$ is itself an eigenfunction of the Laplacian with eigenvalue
$4\pi^2\eigenvalue$, and then appealing to the general results of
Sogge \cite{Sogge} on $L^p$-norms of eigenfunctions. We recall these:
Let
$$
M_{d,p}(E) = \sup_{\Delta f+4\pi^2 \eigenvalue f=0}
\frac{ ||f||_p}{||f||_2} \;.
$$
Then for $p\leq 4$ we have (using $|u|\leq 1$) that $\int u^4\leq
||u||_p^p$ and hence
$$
\int_{\T^d} u^4(x)dx \ll_p \left( \frac{M_{d,p}(E)}{\sqrt{\Ndim}}
\right)^p \;. 
$$

Sogge showed that for eigenfunctions of the Laplacian on any
smooth compact Riemannian manifold, and for for
$p=p_d:=2(d+1)/(d-1)$, one has $M_{d,p}(E)\ll E^{1/2p_d}$. Since
$p_d\leq 4$ for $d\geq 3$,  we have
$$\int_{\T^d} u^4(x)dx\ll \frac{E^{1/2}}{\Ndim^{p_d/2}}\;.
$$

In dimension $d\geq5$ we have $\Ndim\approx E^{\frac d2-1}$ and hence
we find
$$
\int_{\T^d} u^4(x)dx \ll \frac 1{\Ndim^{1+\alpha(d)}}, \quad \alpha(d)
=\frac{2}{d-1}-\frac 1{d-2}
$$
which improves on \eqref{general bound on m4} whenever $d>3$
since $\alpha(d)>0$ for $d>3$.



For the torus in dimension $d\geq 4$, Bourgain
\cite{Bourgain} showed that for $p\geq\frac{2(d+1)}{d-3}$,
$$M_{d,p}(E) \ll E^{\frac{d-2}4-\frac{d}{2p}+\epsilon} , \forall \epsilon>0
$$
which improves on Proposition~\ref{prop:fourth moment}
in dimension $d\geq 7$ (when we may take $p=4$).

\newcommand{\setoutxy}{Sing_{x,y}^{\rm out}}
\newcommand{\setinxy}{Sing_{x,y}^{\rm in}}

\appendix
\section{The intersection of the singular set with codimension one hyperplanes}
\label{app:P_x}

We consider the hyperplane 
$$
\Px^a=\{f\in \eigenspace:\: f(x)=a \}
$$ 
and show that the set of singular functions in $\Px^a$ has measure zero. 
Assume that the set of frequencies $\Lambda$, which is assumed to be
``symmetric'', further satisfies the non-degeneracy
condition~\eqref{condition 1a}, that is:
\begin{equation}\label{condition 1}
\exists \lambda \in \Lambda \mbox{ with } \lambda_1
\neq \pm \lambda_2 \mbox{ and } \lambda_1,\lambda_2 \neq 0 \;.
\end{equation} 
By the symmetry of the set $\Lambda$, condition~\eqref{condition 1} is
equivalent to requiring that for every $i\neq j$, there is $\lambda\in
\Lambda$ with $\lambda_i\neq \pm \lambda_j$ and
$\lambda_i,\lambda_j\neq 0$. 

\begin{proposition}
\label{lem:ssing zero meas Px} 
Assume that $\Lambda$ is symmetric and satisfies the nondegeneracy
condition \eqref{condition 1}. Then for all $x\in \TT^d$, and all $a$, 
the intersection $\Px^a \cap Sing$ has measure zero in $\Px^a$. 
\end{proposition}


In order to prove Proposition \ref{lem:ssing zero meas Px}, we will need
some lemmas.  

Let  $L_{\Lambda}\subset \Z^d$ be the lattice spanned by $\Lambda$. By
Lemma~\ref{lem:Lambda spans Rd}, it is a sublattice of full rank,
hence its dual  $L_{\Lambda}^*$ is still a lattice in $\eigenspace$.   
In \S~\ref{sec:Gaussian ensembles} we defined the set $\B$ by
$$
\B = \{w\in \R^d: \langle \lambda,w \rangle \in \Z\quad \forall
\lambda\in \Lambda \mbox{ or } 
 \langle \lambda,w \rangle \in \frac 12 +\Z\quad \forall
\lambda\in \Lambda \} \;.
$$
Let
$$
\B_x:=\{y\in \T^d: x-y\in  \B \} \;.
$$
Note that if $y\in \B_x$ then for all $f\in \eigenspace$, 
$f(x)=\pm f(y)$ and $\nabla f(x)=\pm
\nabla f(y)$. 

\begin{lemma}\label{lem: no sols}
Suppose that $\Lambda$ is symmetric and satisfies
the nondegeneracy condition~\eqref{condition 1}. 
If $w\notin \B$ then
there are no nonzero solutions $(\vec{c},b',b'')\in \R^d\times
\R\times \R$,  satisfying 
\begin{eqnarray}
\langle \vec{c},\lambda \rangle &=& b''\sin 2\pi \langle w,\lambda \rangle
\label{eqsine}\\
 b' &= & b''\cos 2\pi \langle w,\lambda \rangle \label{eqcos}
\end{eqnarray}
for all $\lambda\in \Lambda$.
\end{lemma}
\begin{proof}
If $b'' = 0$ then $b'=0$ and since $\Lambda$ spans $\R^d$ by Lemma
\ref{lem:Lambda spans Rd}, we find  $\vec{c}=0$. Otherwise, 
from \eqref{eqcos} we find that $\forall \lambda\in \Lambda$
\begin{equation}\label{eq sin pm}
  \sin 2\pi \langle w,\lambda \rangle  = \pm \sqrt{1-(\frac
{b'}{b''})^2} 
\end{equation}
(necessarily $|b'|\leq |b''|$). Set 
$$
\gamma=\sqrt{(b'')^2-(b')^2}\;.
$$

We will show that $\vec{c}=\vec{0}$, which implies that $\sin 2\pi \langle
w,\lambda \rangle  = 0$ for all $\lambda\in \Lambda$,  and thus $\cos2\pi
\langle w,\lambda \rangle = \pm 1$; by \eqref{eqcos}, $\cos2\pi
\langle w,\lambda \rangle $ is constant and so is either $+1$ for all
$\lambda\in \Lambda$ or equals $-1$ for all $\lambda\in \Lambda$, 
hence we will find that $w\in \B$, contradicting our assumption.  

Fix $j=1,\dots, d$ and we wish to see $c_j=0$; by symmetry we may
take $j=1$. 
Find $\lambda\in\Lambda$ satisfying  condition~\eqref{condition 1}. 
Next, replacing $\lambda$ by $-\lambda$ if necessary, we
may assume that
$$
\langle \vec{c},\lambda \rangle = +\gamma
$$
that is
\begin{equation} \label{sum plus l}
\lambda_1 c_1 + \sum_{i\neq 1} c_i \lambda_i = +\gamma \;.
\end{equation}

Let $\hat \lambda=(-\lambda_1,\lambda_2,\dots)\in \Lambda$ be the
result of changing the sign 
of the first coordinate of $\lambda$. 
Then $\langle \hat \lambda,\vec{c}\rangle = \pm \gamma$, that is
\begin{equation} \label{sum pm l}
  -\lambda_1 c_1 + \sum_{i\neq 1} c_i \lambda_i = \pm \gamma \;.
\end{equation}
If the sign is $+$, we compare \eqref{sum pm l} with \eqref{sum
plus l} to deduce that
$$ c_1\lambda_1 =0$$
and since $\lambda_1\neq 0$ we find that  $c_1=0$.

Otherwise, if the sign in \eqref{sum pm l} is $-$, we compare with
\eqref{sum plus l} to find
\begin{equation}\label{eq no 6}
  c_1\lambda_1 = +\gamma \;.
\end{equation}

Repeating the above argument with $\lambda$ replaced by
$(\lambda_2,\lambda_1,\dots)\in \Lambda$ (that is we switch the first and
second coordinates), we find that either $c_1=0$ or else 
%
%
\begin{equation}\label{eq no 9}
  c_1\lambda_2 = +\gamma
\end{equation}
and together with \eqref{eq no 6}  we find that
$$
 c_1\lambda_2 = +\gamma =  c_1\lambda_1 \;.
$$
Since  $\lambda_2\neq \lambda_1$ we find again that $c_1=0$.
\end{proof}

\begin{lemma}\label{lem:psi submer} 
Suppose that $\Lambda$ is symmetric and
satisfies the nondegeneracy  condition~\eqref{condition 1}. 
Then for every $x\in\T^d$, the map
$\Psi_x$  given by
\begin{equation}
\label{eq:tilde psix def}
  \begin{split}
\Psi_x:  (\T^d\backslash \B_x)\times \eigenspace &\to  
\R^d\times \R\times \R  \\
                  (y,f) & \mapsto  (\nabla f(y), f(y), f(x)) 
\end{split}
\end{equation}
is a submersion.
\end{lemma}


\begin{proof}
We wish to show that the derivative 
 $D_{y,f} \Psi_x:\R^d\times \R^{\Ndim} \to
\R^{d+2}$ at the point $(y,f)$ has rank $d+2$. For this it suffices to
show that the $(d+2)\times \Ndim$ matrix 
$\frac{\partial  \Psi_x}{\partial f}$ has rank $d+2$. 
Now 
$$
\frac{\partial  \Psi_x}{\partial f} = 
\bigoplus_{\lambda\in \Lpm} \sqrt{\frac{2}{\Ndim}} 
\begin{pmatrix}
  -2\pi \sin 2\pi \langle \lambda,y \rangle \vec \lambda & 
-2\pi \cos 2\pi \langle \lambda,y \rangle \vec \lambda \\
 \cos 2\pi \langle \lambda,y \rangle & -\sin 2\pi \langle \lambda,y \rangle \\
\cos 2\pi \langle \lambda,x \rangle  & -\sin 2\pi \langle \lambda,x
\rangle 
\end{pmatrix} \;. 
$$
Post-multiplying it by the (block-diagonal) invertible matrix
$$
\bigoplus_{\lambda\in \Lpm}
\sqrt{\frac{\Ndim}{2}} 
\begin{pmatrix}
 -\sin 2\pi \langle \lambda,y \rangle & \cos 2\pi \langle \lambda,y
  \rangle \\ -\cos 2\pi \langle \lambda,y \rangle & 
-\sin 2\pi \langle \lambda,y \rangle
\end{pmatrix}
$$
gives the $(d+2)\times \Ndim$ matrix
$$
\bigoplus_{\lambda\in \Lpm} 
\begin{pmatrix}
2\pi \vec \lambda & \vec 0 \\
0 & 1 \\
\sin 2\pi  \langle \lambda,x-y \rangle  & 
\cos 2\pi  \langle\lambda,x-y \rangle 
\end{pmatrix} \;. 
$$
Thus we want to show that the rank of this matrix  is $d+2$.

For this it suffices to show that the rows are linearly
independent, that is there is no non-trivial solution $(\vec{c},
b',b'')\in \R^{d+2}$ to the system
\begin{eqnarray*}
  \langle \vec{c},\lambda \rangle &=&
b'' \sin 2\pi \langle x-y,\lambda \rangle\\
 b' &=& b'' \cos  2\pi \langle x-y,\lambda \rangle
\end{eqnarray*}
which  by Lemma ~\ref{lem: no sols} this has
no solutions if $x-y\notin \B$, that is if $y\notin \B_x$.
\end{proof}

\begin{proof}[Proof of Proposition~\ref{lem:ssing zero meas Px}]
We will partition $\Px^a \cap Sing$ into two sets: The set
$\setinx$ of those $f$ for which all singular points of the nodal set
of $f$ lie in $\B_x$ (here necessarily $a=0$), and the set $\setoutx$
of those $f$ for 
which there is a singular point of the nodal set  outside $\B_x$.
We will show that each has measure zero.

We first show that $\setinx$ has measure zero. We will in fact see
that it is a linear subspace of codimension $d$ in $\Px^0$. Note
that if $y\in \B_x$ then $f(y) = \pm f(x)$ and $\nabla f(y) = \pm
\nabla f(x)$ and so
$$
\setinx = \{f\in \eigenspace: f(x) = 0, \quad \nabla f(x) = \vec 0 \} \;.
$$
Thus $\setinx$ are the solutions to the linear system of equations
$$ 
f(x)=0, \quad \nabla f(x)=0\;.
$$
The $(d+1)\times |\Lambda|$ matrix of this system is
$$
\bigoplus_{\lambda\in \Lpm}
\begin{pmatrix}
  -2\pi \sin 2\pi \langle \lambda,x \rangle \vec \lambda& -2\pi \cos2\pi \langle \lambda,x \rangle \vec \lambda \\
  \cos2\pi \langle \lambda,x \rangle& -\sin 2\pi \langle \lambda,x
\rangle\end{pmatrix} 
$$
which as we have seen in the proof of Lemma \ref{lem:Sing meas 0}
has rank $d+1$, and thus $\setinx \subset \Px^0$ has codimension $d$
in $\Px^0$.

We now turn to $\setoutx$. Let $\pi_{\eigenspace}:\T^d\times
\eigenspace \to \eigenspace$  
be the projection on the second factor; then by the definition 
\eqref{eq:tilde psix def} of $\Psi_x$, 
$$
\pi_{\eigenspace}(\Psi_x^{-1}(\vec 0,0,a)) = \setoutx \;.
$$

Lemma \ref{lem:psi submer} shows, in particular, that 
$(\vec 0,0,a)$ is a regular value of $\Psi_x$, so that 
$\Psi_x^{-1}(\vec 0,0,a)$ is a submanifold of $\T^d\times \eigenspace$ of
codimension $d+2$, that is $\Psi_x^{-1}(\vec 0,0,a)\subset
\T^d\times \Px^a$ has dimension $|\Lambda| - 2$. Therefore $\setoutx
= \pi_{\eigenspace}(\Psi_x^{-1}(\vec 0,0,a)) \subset \Px^a$ 
has dimension
at most $|\Lambda|-2$ in the $(|\Lambda|-1)$-dimensional space
$\Px^a$ and hence has measure zero.
\end{proof}

\section{The intersection of the singular set with codimension two
hyperplanes} 
For $a=(a_1,a_2) \in \R^2$, let 
$$
\Pxy^a = \{f\in \eigenspace: f(x)=a_1, f(y)=a_2 \}\;.
$$
If $x-y\notin \B$ then this is an affine hyperplane
of codimension two in $\eigenspace$.  
If $x-y\in \B$ then this is either empty or a
hyperplane of codimension one in $\eigenspace$.

%

\begin{proposition}
\label{lem:SSing in Pxy is zero} 
For $d\geq 2$, for any symmetric set of frequencies $\Lambda$
satisfying the non-degeneracy condition~\eqref{condition 1}, 
there is a set of measure zero $S=S_\Lambda \subset \TT^d$ so that for
$x-y\neq S$, the intersection  $\Pxy^a\cap Sing $ has measure
zero in $\Pxy^a$. 
\end{proposition}

The proof of Proposition~\ref{lem:SSing in Pxy is zero} 
follows along the lines of Proposition~\ref{lem:ssing zero meas Px}, 
proving that the codimension is $\ge 1$. 
We will need a lemma about the nonexistence of solutions to certain
systems of equations:

\begin{lemma}\label{lem:lin ind meas d>2} 
Let $d\geq 2$. Then for any symmetric set of frequencies $\Lambda$ 
satisfying the non-degeneracy condition~\eqref{condition 1},  there is
a set $S\subset 
\TT^d$ of measure zero so that if $x-y\notin S$ then there do not exist  
 $z\in \T^d$, numbers $b_1,b_2\neq 0$
and $b_3$ and $\vec{c}\in \R^d$, which
satisfy
\begin{equation}
\label{eq:dep cond} 
b_3+ i\langle \vec{c},\,\lambda\rangle = 
b_1 e^{2\pi i \langle \lambda,\,x-z\rangle} + 
b_2 e^{2\pi i \langle \lambda,\, y-z\rangle} 
\end{equation}
for every $\lambda\in\Lambda$.
\end{lemma}
\begin{proof}
We choose $\lambda\in \Lambda$ satisfying 
condition~\eqref{condition 1}, that is $\lambda_1,\lambda_2\neq0$ and
$\lambda_1\neq \pm\lambda_2$.  
Taking the  norm-square of \eqref{eq:dep cond}, we have
\begin{equation*}
b_3^2 +\langle \vec{c},\lambda \rangle^2 = b_1^2 +b_2^2 +2b_1 b_2 \cos2\pi
\langle \lambda, x-y \rangle \;.
\end{equation*}
Now repeat this with $\lambda$ replaced by  
$$
\lambda^\epsilon:=(\epsilon_1\lambda_1,\epsilon_2\lambda_2
,\dots, \epsilon_d \lambda_d) 
$$
and sum the resulting equalities over all $\epsilon\in \{\pm 1\}^d$,
each weighted by  
$$
\chi_{1,2}(\epsilon) = 
\epsilon_1 \epsilon_2 \;.
$$
This gives 
\begin{multline*}
\sum_{\epsilon\in \{\pm 1\}^d} \chi_{1,2}(\epsilon) \left( 
b_3^2 +\langle \vec{c},\lambda^\epsilon\rangle^2 
\right) \\
= \sum_{\epsilon\in \{\pm 1\}^d} \chi_{1,2}(\epsilon) 
\left( b_1^2 +b_2^2 +2b_1 b_2 \cos2\pi \langle \lambda^\epsilon , x-y
\rangle \right) \;.
\end{multline*}
Now use 
$$\sum_{\epsilon\in \{\pm 1\}^d} \chi_{1,2}(\epsilon)=0$$
to get 
\begin{equation*}
\sum_{\epsilon\in \{\pm 1\}^d} \chi_{1,2}(\epsilon)\langle \vec{c},\lambda^\epsilon
\rangle^2  = 2b_1b_2 \sum_{\epsilon\in \{\pm 1\}^d}\chi_{1,2}(\epsilon) 
\cos 2\pi\langle \lambda^\epsilon , x-y \rangle \;.
\end{equation*}
Expand  
$$
\langle \vec{c},\lambda^\epsilon \rangle^2 = \sum_{j,k=1}^d
\lambda_j\lambda_k c_j c_k \epsilon_j \epsilon_k
$$
and use
$$
\sum_{\epsilon\in \{\pm 1\}^d}\chi_{1,2}(\epsilon) \epsilon_j\epsilon_k = 
\begin{cases}
2^d,& (j,k)=(1,2) \mbox{ or } (2,1) \\ 0& \mbox{ otherwise }
\end{cases}
$$
to get
$$ 
\sum_{\epsilon\in \{\pm 1\}^d}\chi_{1,2}(\epsilon) \langle \vec{c}, \lambda^\epsilon
\rangle^2 = 2^{d+1} c_1 c_2 \lambda_1 \lambda_2 \;.
$$
Thus we find
\begin{equation}\label{first try}
2^{d+1} c_1 c_2 \lambda_1 \lambda_2 = 2b_1 b_2\sum_{\epsilon\in \{\pm 1\}^d}
\chi_{1,2}(\epsilon) \cos 2\pi\langle \lambda^\epsilon,x-y \rangle \;.
\end{equation}

We repeat the argument with $\lambda$ replaced by 
$$\tilde\lambda = (\lambda_2,\lambda_1,\lambda_3,\dots, \lambda_d)$$
that is we have permuted the first and second coordinates of
$\lambda$. Then we get
\begin{equation}\label{second try}
2^{d+1} c_1 c_2 \lambda_2 \lambda_1 = 2b_1 b_2\sum_{\epsilon\in \{\pm 1\}^d}
\chi_{1,2}(\epsilon) \cos 2\pi\langle \tilde\lambda^\epsilon,x-y
\rangle \;. 
\end{equation}
Comparing \eqref{first try} with \eqref{second try} and dividing by 
$2b_1b_2$ (which is nonzero by assumption), we get
\begin{equation}\label{identity for cos}
\sum_{\epsilon\in \{\pm 1\}^d} \chi_{1,2}(\epsilon)\cos 2\pi
\langle \lambda^\epsilon,x-y \rangle = 
\sum_{\epsilon\in \{\pm 1\}^d} \chi_{1,2}(\epsilon)\cos 2\pi
\langle \tilde\lambda^\epsilon,x-y \rangle \;.
\end{equation}
Writing 
$$
\cos 2\pi\langle \lambda^\epsilon,x-y \rangle  = 
\frac {\exp 2\pi i \langle \lambda^\epsilon,x-y \rangle + \exp 2\pi i
\langle \lambda^{-\epsilon},x-y \rangle } 2
$$  
and noting that 
$\chi_{1,2}(-\epsilon)=\chi_{1,2}(\epsilon)=\epsilon_1\epsilon_2$, 
we may rewrite
\eqref{identity for cos} as
\begin{equation}\label{identity 2 for cos}
\sum_{\epsilon\in \{\pm 1\}^d}\epsilon_1\epsilon_2 \exp 2\pi i \langle
\lambda^\epsilon,x-y \rangle = 
\sum_{\epsilon\in \{\pm 1\}^d}\epsilon_1\epsilon_2 \exp 2\pi i \langle
\tilde \lambda^\epsilon,x-y \rangle \;. 
\end{equation}
If we 
use the identity
$$
\sum_{\epsilon_3,\dots,\epsilon_d = \pm 1} \exp 2\pi i \sum_{j=3}^d
\epsilon_j \lambda_j(x_j-y_j)
= 2^{d-2}\prod_{j=3}^d \cos2\pi \lambda_j(x_j-y_j)
$$
and some simple trigonometric identities, 
then \eqref{identity 2 for cos} becomes 
\begin{multline*}
2^{d-1} \sin 2\pi \lambda_1(x_1-y_1)\sin 2\pi \lambda_2(x_2-y_2) 
\prod_{j=3}^d \cos2\pi \lambda_j(x_j-y_j)
 \\  
= 2^{d-1}\sin 2\pi \lambda_2(x_1-y_1)\sin 2\pi \lambda_1(x_2-y_2)
 \prod_{j=3}^d \cos2\pi \lambda_j(x_j-y_j) \;.
\end{multline*}
This forces either 
$$
\sin 2\pi \lambda_1(x_1-y_1)\sin 2\pi \lambda_2(x_2-y_2)=
\sin 2\pi \lambda_2(x_1-y_1)\sin 2\pi \lambda_1(x_2-y_2)\;,
$$
which is a measure zero condition on $x-y$ since we assume that
$\lambda_1,\lambda_2\neq 0$ and $\lambda_1\neq \pm \lambda_2$,   
or else $d\geq 3$ and there is some  $j\neq 1,2$  with $\lambda_j\neq
0$  for which $\cos 2\pi \lambda_j(x_j-y_j)=0$, 
which is again a measure zero condition on $x-y$. 
\end{proof}

As before, we denote by $\B_x=x+\B$. 
For $x,y\in \TT^d$, $x-y \notin \B$, consider the
map
\begin{equation}\label{defn of Psixy}
\begin{split}
\Psi_{x,y}:\TT^d\backslash (\B_x\cup \B_y) \times \eigenspace &\to
\R^{d+3} \\
(z,f)&\mapsto (\nabla f(z), f(z),f(x), f(y))
\end{split}
\end{equation}

\begin{lemma}\label{lem:submersion 2}
Suppose that $\Lambda$ is symmetric and satisfies the non-degeneracy 
condition~\eqref{condition 1}. 
Then there is a set $S=S_\Lambda\subset \TT^d$ of measure zero so that if
$x-y\notin S$, then $\Psi_{x,y}$ is a submersion.
\end{lemma}
\begin{proof}
We wish to show that the derivative 
 $D_{z,f} \Psi_{x,y}:\R^d\times \R^{\Ndim} \to
\R^{d+3}$ at the point $(z,f)$ has rank $d+3$. For this it suffices to
show that the $(d+3)\times \Ndim$ matrix 
$\frac{\partial  \Psi_{x,y}}{\partial f}$ has rank $d+3$. 
Now 
$$
\frac{\partial  \Psi_{x,y}}{\partial f} = 
\bigoplus_{\lambda\in \Lpm} \sqrt{\frac{2}{\Ndim}} 
\begin{pmatrix}
  -2\pi \sin 2\pi \langle \lambda,z \rangle \vec \lambda & 
-2\pi \cos 2\pi \langle \lambda,z \rangle \vec \lambda \\
 \cos 2\pi \langle \lambda,z \rangle & -\sin 2\pi \langle \lambda,z \rangle \\
\cos 2\pi \langle \lambda,x \rangle  & -\sin 2\pi \langle \lambda,x
\rangle \\
\cos 2\pi \langle \lambda,y \rangle  & -\sin 2\pi \langle \lambda,y
\rangle 
\end{pmatrix} \;. 
$$
Post-multiplying it by the (block-diagonal) invertible matrix
$$
\bigoplus_{\lambda\in \Lpm}
\sqrt{\frac{\Ndim}{2}} 
\begin{pmatrix}
 -\sin 2\pi \langle \lambda,z \rangle & \cos 2\pi \langle \lambda,z
  \rangle \\ -\cos 2\pi \langle \lambda,z \rangle & 
-\sin 2\pi \langle \lambda,z \rangle
\end{pmatrix}
$$
gives the $(d+3)\times \Ndim$ matrix
$$
\bigoplus_{\lambda\in \Lpm} 
\begin{pmatrix}
2\pi \vec \lambda & \vec 0 \\
0 & 1 \\
\sin 2\pi  \langle \lambda,x-z \rangle  & 
\cos 2\pi  \langle\lambda,x-z \rangle \\
\sin 2\pi  \langle \lambda,y-z \rangle  & 
\cos 2\pi  \langle\lambda,y-z \rangle 
\end{pmatrix} \;. 
$$
Thus we want to show that the rank of this matrix  is $d+3$, 
that is that the rows are linearly independent, i.e. 
that is there is no non-trivial solution 
$(\vec{c},b_1,b_2,b_3)\in \R^{d+3}$ so that 
\begin{equation*}
\begin{split}
\langle \vec{c},\, \lambda \rangle = b_1 \sin{2\pi \langle \lambda,\,
x-z
\rangle} + b_2 \sin{2\pi \langle \lambda,\, y-z \rangle} \\
b_3 = b_1 \cos{2\pi \langle \lambda,\, x-z \rangle} + b_2 \cos{2\pi
\langle \lambda,\, y-z \rangle},
\end{split}
\end{equation*}
for all $\lambda\in\Lambda$. 
We may write the system in a
complex form as
\begin{equation*}
 b_3+i\langle \vec{c},\,\lambda \rangle =b_1e^{2\pi i
\langle \lambda,\, x-z\rangle }+b_2 e^{2\pi i \langle \lambda,\, y-z\rangle}.
\end{equation*}

If either of $b_1,\, b_2$ is zero, we
are in the same situation as in Lemma~\ref{lem: no sols} 
and so we deduce that either $x-z\in \B$ or 
$y-z\in \B*$, contradicting our assumption that
$z\notin \B_x \cup \B_y$. 
If both $b_1,b_2\neq 0$, then Lemma~\ref{lem:lin ind meas d>2} implies
the result of Proposition~\ref{lem:SSing in Pxy is zero}. 
\end{proof}

\noindent{\bf Proof of Proposition~\ref{lem:SSing in Pxy is zero}:} 
Given the measure zero set $S$ of 
Lemma~\ref{lem:lin ind meas d>2}, and $x,y\in\TT^d$ with $x-y\notin
S$, we write the set of singular elements in 
$\Pxy^a$ as a union of two subsets each of which we will show to have
measure zero: 
$$ \Pxy^a\cap Sing = \setinxy \cup \setoutxy$$
where:

i)  $\setinxy$ consists of those $f\in \Pxy^a$ for which all  
singular points of the nodal set (that is $z$ so that $f(z)=0$, 
$\nabla f(z)=0$) lie in $\B_x \cup \B_y$. If $z\in \B_x$ then 
$f(x)=\pm f(z)$ and $\nabla f(x)=\pm \nabla f(z)$ so either $f(x)=0$,  
$\nabla f(x)=0$ or the same with $y$ replacing $x$. If both
$a_1,a_2\neq 0$ then $\setinxy=\emptyset$, and in any case 
we will see that $\setinxy$ has measure zero in $\Pxy^a$: 
Indeed, as we saw in
Lemma~\ref{lem:Sing meas 0}, for every $x\in \TT^d$, the linear space
$$
 \{f\in \eigenspace: f(x) = 0, \nabla f(x)=0\}
$$
has codimension $d+1$ in $\eigenspace$. Since 
$\Pxy^a$ has codimension $2$ in $\eigenspace$, we find that 
$\setinxy$ is a union of two affine hyperplanes of codimension at
least $d-1\geq 1$ in $\Pxy^a$ (recall $d\geq 2$), and therefore has
measure zero in $\Pxy^a$.

ii) $\setoutxy$ consists of $f\in \Pxy^a$ for which there is a singular 
point $z$ of the nodal set outside of $\B_x\cup \B_y$. Thus in the
notation of \eqref{defn of Psixy}, 
$$ \setoutxy = \pi_\eigenspace \circ \Psi_{x,y}^{-1}(\vec{0},0,a)$$  
where $\pi_\eigenspace:\TT^d\times \eigenspace \to \eigenspace$ is the
projection onto the second factor. Since $x-y\notin S$, we may use
Lemma~\ref{lem:submersion 2} to deduce that
$\Psi_{x,y}^{-1}(\vec{0},0,a)$ is a 
submanifold of  $\TT^d \times \eigenspace$ of codimension $d+3$, hence
its projection $\pi_\eigenspace \circ \Psi_{x,y}^{-1}(\vec{0},0,a)$
has codimension at least $3$ in 
$\eigenspace$ and hence codimension at least one in $\Pxy^a$. 
Thus $\Pxy^a\cap Sing$ has measure zero in $\Pxy$, in fact has
codimension at least one.
\qed


\begin{thebibliography}{99}


\bibitem{Berard}
P.~B\'erard, {\em Volume des ensembles nodaux des fonctions propres du
  laplacien}.   Bony-Sjostrand-Meyer seminar, 1984--1985,
  Exp. No. 14 , 10 pp., \'Ecole Polytech., Palaiseau, 1985.

\bibitem{Berry 1977}
M.~V.~Berry, {\em Regular and irregular semiclassical wavefunctions}.
J. Phys. A  {\bf 10}  (1977), no. 12,        2083--2091.

\bibitem{Berry 2002}
M.~V.~Berry,  {\em Statistics of nodal lines and points in
chaotic quantum billiards: perimeter corrections, fluctuations,
curvature} J.Phys.A {\bf 35} (2002), 3025-3038.

\bibitem{BR}
M.~Borovoi and Z.~Rudnick, {\em Hardy-Littlewood varieties and semisimple
groups}, Inventiones Math {\bf 119} , 37--66 (1995).

\bibitem{Bourgain}
J.~Bourgain, {\em Eigenfunction bounds for the Laplacian on the
$n$-torus}, Internat. Math. Res. Notices 1993, no. 3, 61--66.

\bibitem{Cilleruelo}
J.~Cilleruelo, {\em The distribution of the lattice points on
circles}. J. Number Theory {\bf 43} (1993), no. 2, 198--202.

\bibitem{Davenport}
H.~Davenport, {\em Analytic methods for Diophantine equations and
Diophantine inequalities}. Second edition. With a foreword by
R. C. Vaughan, D. R. Heath-Brown and D. E. Freeman. Edited and
prepared for publication by T. D. Browning. Cambridge Mathematical
Library. Cambridge University Press, Cambridge, 2005.

\bibitem{EH}
P.~Erd\"os and R.~R.~Hall, {\em On the angular distribution of Gaussian
integers with fixed norm}, Discrete Math., {\bf 200} (1999), pp. 87--94.
(Paul Erd\"os memorial collection).

\bibitem{FKW}
L.~Fainsilber, P.~Kurlberg and B.~Wennberg, {\em Lattice points on
circles and discrete velocity models for the Boltzmann equation}. SIAM
J. Math. Anal. {\bf 37} (2006), no. 6, 1903--1922.

\bibitem{GS1}
I.~M.~Gelfand and G.~E.~Shilov,  {\em Generalized
functions}. Vol. 1. Properties and operations. Translated from the
Russian by Eugene Saletan. Academic Press [Harcourt Brace Jovanovich,
Publishers], New York-London, 1964 [1977].

\bibitem{Kac}
M.~Kac, {\em On the average number of real roots of a random algebraic
 equation}. Bull. Amer. Math. Soc. {\bf 49},
(1943), 314--320. Correction, ibid.  {\bf 49}, (1943) 938.

\bibitem{KK}
I.~K\'atai and I.~K\"ornyei, {\em On the distribution of lattice points on
circles}, Ann. Univ. Sci. Budapest. Eotvos Sect. Math., {\bf 19} (1977),
pp. 87--91.

\bibitem{LH1}
M.~S.~Longuet-Higgins, {\em The statistical analysis of a random,
moving surface}. Philos. Trans. Roy. Soc. London Ser. A. {\bf 249} (1957),
321--387.

\bibitem{LH2}
M~.S.~Longuet-Higgins, {\em  Statistical properties of an isotropic
random surface}. Philos. Trans. Roy. Soc. London. Ser. A. 250 (1957),
157--174.

\bibitem{Neuheisel}
J.~Neuheisel, {\em The asymptotic distribution of nodal sets on
  spheres}, Johns Hopkins Ph.D. thesis (2000).

\bibitem{Palamodov}
V.~P.~Palamodov,  {\em Distributions and harmonic analysis}, in
Commutative harmonic analysis, III (Havin and N.K. Nikol'skij, eds.),
1--127, 261--266, Encyclopaedia Math. Sci., 72, Springer, Berlin, 1995.

\bibitem{Pommerenke}
C.~Pommerenke,
{\em \"Uber die Gleichverteilung von Gitterpunkten auf
$m$-dimensionalen Ellipsoiden}, Acta Arith. {\bf 5} 1959 227--257.
Erratum in  Acta Arith. {\bf 7} 1961/1962 279.


\bibitem{RW}
Z.~Rudnick and I.~Wigman, {\em On the volume of nodal sets for
eigenfunctions of the Laplacian on the torus}, in preparation. 

\bibitem{Sogge}
C.~D.~Sogge, {\em Fourier integrals in classical analysis}. 
Cambridge Tracts in Mathematics, 105.
Cambridge University Press, Cambridge, 1993.


\bibitem{ZelditchRMT} 
S.~Zelditch, {\em A random matrix model for quantum mixing}.  
Internat. Math. Res. Notices (1996), Issue 3, Pages 115-137.  




 

\bibitem{Zygmund}
A.~Zygmund, {\em On Fourier coefficients and transforms of functions of
two variables}, Studia Math. {\bf 50} (1974), 189--201.




\end{thebibliography}
\end{document}